\documentclass[a4paper,11pt]{article}
\usepackage{jcappub}

\newcommand{\Lya}{Lyman-$\alpha$}

\title{More accurate simulations with separate initial conditions for baryons and dark matter}
\author[a,1]{Simeon Bird\note{Corresponding author}}
\author[b]{Yu Feng}
\author[c]{Christian Pedersen}
\author[c]{Andreu Font-Ribera}
\affiliation[a]{Department of Physics \& Astronomy, University of California Riverside,\\ Riverside, CA 92521, USA}
\affiliation[b]{Berkeley Center for Cosmological Physics, University of California Berkeley, \\Berkeley, CA 94720, USA}
\affiliation[c]{Department of Physics \& Astronomy, University College London,\\Gower Street, London WC1E 6BT, UK}

\emailAdd{sbird@ucr.edu}
\emailAdd{yfeng1@berkeley.edu}
\emailAdd{christian.pedersen.17@ucl.ac.uk}
\emailAdd{a.font@ucl.ac.uk}

\abstract{
We revisit techniques for performing cosmological simulations with both baryons and cold dark matter when each fluid has different initial conditions, as is the case at the end of the radiation era. Most simulations do not reproduce the linear prediction for the difference between the cold dark matter and baryon perturbations. We show that this is due to the common use of offset regular grids when setting up the particle initial conditions. The desired linear evolution can be obtained without any loss of simulation resolution by using a Lagrangian glass for the baryon particles. We further show that the difference between cold dark matter and baryons may affect predictions for the Lyman-$\alpha$ forest flux power spectrum at the $5\%$ level, potentially impacting current cosmological constraints.
}

\begin{document}

\maketitle

\section{Introduction}

Cosmological N-body simulations are a well-established technique for understanding non-linear structure formation. A common technique for N-body simulations is model a single fluid, corresponding to a combination of cold dark matter (CDM) and baryons, under gravity, under the approximation that these two components trace each other. Hydrodynamic simulations follow baryons using a separate particle species, but frequently use the same initial transfer function for each species \cite[e.g.][]{Emberson:2018}.

However, at early times the baryons couple to radiation, while the CDM does not. This both induces the baryon acoustic oscillation (BAO) peak  and reduces the clustering of baryonic matter on scales less than the comoving horizon scale at recombination. Structure formation reduces this relative difference between baryons and CDM, as each species couples equally to the gravitational potential. By $z=0$ the power spectra of baryons and CDM differ by less than $1\%$. Nevertheless, at higher redshift they differ substantially, by up to $8\%$ at $z=10$ and $3\%$ at $z=2$ for $ k > 0.02$ k/Mpc.

As observational probes of the Universe reach ever higher redshift, these differences may become more relevant. However, naively modelling separate transfer functions for each species in simulations does not reproduce a key prediction of linear theory, the evolution of the offset in power between the CDM and baryon species \cite{OLeary:2012, Angulo:2013}. We show that using two offset grids for the baryons and CDM causes the relative differences in power to be over-estimated due to a spurious growing mode. Even with separate species, simulations correctly evolve the total matter spectrum and thus may be used to produce, for example, accurate galaxy halo catalogues. However, some cosmological probes, such as the \Lya~forest \cite{PD2013}, $21$-cm power spectrum \cite{Naoz:2005}, the earliest structures \cite{Naoz:2007,Naoz:2012,Popa:2016} or even globular clusters \cite{Chiou:2019} are sensitive specifically to gas at high redshift. An accurate cosmological analysis of these probes should correctly reproduce the linear theory evolution of each species.

In this paper we discuss several methods of achieving this. Our preferred resolution is to initialize the (pre-displacement) baryons using a Lagrangian glass rather than a grid offset from the CDM. A Lagrangian glass initializes particles using a homogeneous quasi-random tessellation of the mass density derived by minimizing the gravitational potential \cite{White:1994}. Other solutions are over-sampling the CDM relative to baryons by a factor of $\Omega_\mathrm{CDM}/\Omega_\mathrm{b}$ (so that particle of each type has the same mass), or using an adaptive gravitational softening for gas particles. We evaluate and compare each method. We also discuss how linear perturbation theory can partially clarify the discrepancy that occurs with two offset grids. Finally, we show that using species-dependent transfer functions affects the \Lya~forest at the $5-10\%$ level on scales measured by the BOSS survey \cite{PD2013}.

%We also consider other methods for resolving the particle discrepancy, showing that it is not due to force inaccuracy.

%The noise inherent in a glass distribution is reduced by using a glass only for the baryons, not for the CDM. To further prevent chance juxtapositions of CDM and baryon particles we perform reverse-gravity timesteps on the combined particle distributions.

Several earlier works have treated aspects of the same question. Refs~\cite{OLeary:2012, Angulo:2013} showed that linear theory can be reproduced using adaptive softening lengths. With adaptive softening lengths, the gravitational softening of the baryon particles is set proportional to their SPH smoothing length. This effectively disables the short range gravitational force at high redshift, when the Universe is homogeneous and SPH smoothing lengths are of order the mean inter-particle spacing.

This procedure is sometimes justified on the basis of the Jeans length of a pressure-supported cloud \cite{Fire2:2018} We emphasize that on cosmological scales this justification does not apply as pressure forces are negligible on these scales. The specific adaptive softening length chosen, the SPH smoothing length, thus cannot be justified on the basis of hydrodynamics. The coupling effect is purely gravitational, and persists even if (as in our simulations and those of Ref.~\cite{Angulo:2013}) hydrodynamical pressure forces are disabled. In practice therefore, adaptive softening has an effect similar to uniformly increasing the softening only at high redshifts where the problem is most acute. Since the small-scale gravitational force is disabled at high redshift, adaptive softening strongly suppresses the small-scale clustering of the baryons in low density regions outside galaxies. This may make it unsuitable for some applications, in particular simulations of the \Lya~forest. Nevertheless, adaptive softening is widely used in recent simulations \cite[e.g][]{Paco:2018}.\footnote{Ref.~\cite{Valkenburg:2017} instead achieved the same effect by setting the cutoff distance for the small-scale force to a small fraction of the PM grid size.}

An alternative resolution to this problem was proposed by Ref.~\cite{Yoshida:2003}. They showed that the desired relative power between baryons and cold dark matter could be achieved by initializing the particles using two independent Lagrangian glasses, rather than the traditional offset grids. Several of our results are implicit in Ref.~\cite{Yoshida:2003}, and in related work discussing the effect of the small-scale supersonic baryon streaming velocity \cite{Naoz:2011}. However, they did not consider large scales and thus did not include the effect of species and scale dependent velocity transfer functions as we do here. In particular, on the scales we consider the relative velocity baryon-CDM streaming velocity \cite{TkHirata:2010} is naturally included as part of the initial velocity transfer function.

% This means that each particle species has roughly the same mass and thus the inaccuracies resulting from differing mass particles are removed.

Other possibilities for the initial particle distribution have also been proposed. Ref.~\cite{Hansen:2007} used a quaquaversal tiling of three dimensional space, which we do not consider as it restricts the total number of particles to a power of two. Ref.~\cite{Liao:2018} used a capacity constrained Voronoi tessellation, which preserves most of the good properties of a Lagrangian glass. We do not use it here only because it is not obvious how to generalize it to two fluids. Both of the above are examples of other quasi-random, homogeneous tessellations of space, like a glass file. Finally, Ref.~\cite{Zennaro:2017} avoid the problem (but only at $z=0$) by ``backscaling'', where each species is scaled separately by the requisite factor so that the final simulation output exactly matches the linear theory prediction.

The rest of this paper proceeds as follows: in Section~\ref{sec:methods} we discuss in detail our techniques for initialising cosmological simulations. In Section~\ref{sec:glass} we review the formation of a glass file. Section~\ref{sec:particles} explains how we set the initial displacements and velocities of the particles. Section~\ref{sec:simulations} describes our example simulation suite. Our results for the matter power spectrum are described in Section~\ref{sec:results}. Section~\ref{sec:offsetgrid} reproduces the existing discrepancy with linear theory. Section~\ref{sec:halfglass} shows that there is good agreement with linear theory for our preferred strategy of a Lagrangian glass for baryonic particles. Section~\ref{sec:otherstrat} examines other successful strategies, comparing their strengths and weaknesses. Section~\ref{sec:explanation} discusses and explains our results, noting the mechanism by which each successful simulation strategy works.
Section~\ref{sec:lymanalpha} shows the effect on the \Lya~forest of a species-dependent transfer function.
Finally, Section~\ref{sec:conclude} concludes.

\section{Methods}
\label{sec:methods}

In this Section we describe our methods for initializing cosmological simulations, as well as the simulations we run.
The basic algorithm of our preferred setup is similar to that presented in Ref~\cite{Yoshida:2003} and is as follows:
\begin{enumerate}
 \item Produce a (homogeneous) distribution of CDM particles arranged in a regular grid.
 \item Produce a (homogeneous) distribution of gas particles from a Lagrangian glass.
 \item Evolve the mixed distribution of both types of particles using a reversed gravitational force, as when generating a Lagrangian glass. This avoids random close juxtapositions of CDM and baryon particles.
 \item Set the velocity of the baryon and CDM particles using the scale-dependent velocity transfer function specific to each species.
  \item Displace the baryon and CDM particles using the displacement transfer function specific to each species.
\end{enumerate}

Section~\ref{sec:glass} describes Steps 1-3, which generate a homogeneous distribution of particles with minimal residual power proportional to $k^4$. These are used as our pre-displacement initial sampling of matter. In Steps 4-5, this homogeneous particle distribution is displaced so as to be a Gaussian random field with the velocity and position power spectrum given by cosmological perturbation theory. These steps are described in Section~\ref{sec:particles}.

\subsection{Grid and Glass Particle Distribution Generation}
\label{sec:glass}

We generate glass particle distributions following Ref.~\cite{White:1994}. A gravitationally interacting N-particle system is evolved with the sign of gravity reversed, so that each particle feels a repulsive gravitational force from every other particle. We have implemented this gravitational force evolution directly into our initial conditions code and it thus uses the particle-mesh algorithm from our N-body code, MP-Gadget. The initial particle distribution before glass generation is a regular grid with the positions of each particle randomized over three mean inter-particle spacings. A symplectic leap-frog is run with a kick-drift-force-kick arrangement for $14$ timesteps\footnote{This algorithm is chosen to allow code-sharing with MP-Gadget's PM code, but symplecticity is likely not required.}, the smallest number of steps that produced a small-scale noise power spectrum in good agreement with the expected $k^4$ shape. The velocity kick includes a damping term proportional to the velocity to avoid particle positions oscillating around the minimum. The velocity produced by the glass evolution is zeroed after the glass generation is complete.

This procedure can be applied separately to both CDM and baryons, generating two uniform homogeneous gas distributions.
Although the distribution of each individual particle species is homogeneous, there is no guarantee that this property also applies to the total particle distribution. The original benefit of a glass, low residual force between all particles, does not hold for two combined glasses \cite{Yoshida:2003} as particles of one species may occasionally, by chance, be initialized extremely close to a particle of a different species. This causes two types of problems: the first is that close pairs of particles causes the computation of the gravitational force to be computationally expensive. The second is that there will be an initial over-density around the particle pair, which may source numerical structure growth.

We investigated generating a single coherent glass file containing both species of particle. In this glass file both particle species are placed at random and then evolved with reversed gravity to a single, coherent, homogeneous distribution. However, this contained no mechanism for keeping the distribution of each species homogeneous, and thus produced a density map which had over-densities of baryons exactly compensated by under-densities of cold dark matter. This in turn led to a very noisy power spectrum estimation as not all modes were fully sampled. Furthermore, in a realistic hydrodynamical simulation it would have caused, for example, the formation of dark matter halos containing no baryons.

Ref.~\cite{Yoshida:2003} proposed a resolution to reduce chance alignments of two particle species. After generating two separate homogeneous glass files, they ran a single reversed-gravity timestep on the combined particle distribution. This ensures that particles which are exceptionally close by chance are disrupted, but does not disrupt the overall glass distribution of the simulation. We perform $14$ reversed gravity timesteps on the combined particle distribution, so that the overall particle distribution is homogeneous. As each species is roughly homogeneous, the residual gravitational force for each species is small and so for most particles the distribution is not stongly perturbed. The number of reversed gravity steps is likely excessive, but once homogeneity is achieved more steps have no effect. We thus use $14$ timesteps as a careful default for larger simulations than our tests.

Lacking the symmetry of a regular cubic grid, glass files produce spurious noise proportional to $k^4$ \cite{Peebles:1993}. Since our use of a glass distribution is solely designed to reproduce the linear theory offset between baryons and cold dark matter, we can reduce the glass noise by using a regular grid for the cold dark matter.\footnote{A grid also has an associated Fourier-space systematic: a broad peak at the grid scale. As with glass noise, this is eventually dominated by structure growth.} Glass noise is still present, but at a reduced level because baryons are only $\sim 1/6$ of the total matter. We also show that in practice hydrodynamical forces suppress the small scale power in baryons, reducing glass noise still further. For realistic particle loads we found that the noise from the glass quickly became subdominant to the physical power spectrum from structure formation, but we use a regular grid for the CDM nevertheless.

% We also considered using $4$ timesteps rather than $14$ timesteps for the reversed gravity evolution of the combined CDM grid and baryonic glass. We found minimal difference in the final power spectra, but the extra timesteps did moderately reduce the number of close particle pairs in the simulation with some improvements in computational time, hence we chose to use $14$ timesteps for the combined particle distribution. \YF{this would show up as a deviation from k ** 4 power. I think we can add a figure showing the glass power vs k**4 with 4 and 14 steps. e.g. a polished version of cell 51 in the notebook. 3-3-2 is the configure we use.}

\subsection{Initialization of Particle Velocities and Displacements}
\label{sec:particles}

Key to reproducing linear theory is the correct initialization of particle velocities and displacements. Since there are several approximations current in the literature, especially for the velocity transfer functions, we here review the linear results.

We initialize both displacements and velocities directly using the transfer functions from the linear Boltzmann code CLASS \cite{CLASS}, corresponding to the solutions of the equations of linear cosmological perturbation theory. We apply a species-independent boost to the velocities so that the simulation takes place in a constant-time Newtonian frame. This is because the CLASS transfer functions are generated in synchronous gauge, a frame comoving with the cold dark matter, and we must convert to the Newtonian frame of the simulation code. Following the notation of Ref.~\cite{Ma:1995}, we set:
\begin{align}
 v_\mathrm{CDM} = \dot{\delta}_\mathrm{CDM}  &= - \frac{\dot{h}}{2} \\
 v_\mathrm{b} = \dot{\delta}_\mathrm{b}  &= - \theta_\mathrm{b} - \frac{\dot{h}}{2}
\end{align}
where $\dot{h}$ and $\theta_\mathrm{b}$ are set using the relevant columns in the CLASS transfer functions. $h$ is the synchronous gauge density perturbation, and $\theta_\mathrm{b}$ the baryon velocity dispersion.
Particle displacements are set using the columns for $\delta_\mathrm{CDM, b}$. Note that each of these quantities is fully scale-dependent, so that the scale-dependence of the velocity perturbations is automatically included without having to numerically differentiate the growth factor\footnote{See: \protect\url{https://github.com/MP-Gadget/MP-Gadget/blob/FirstDOI/libgenic/power.c\#L204}.}.

We do not use the common Zel'dovich approximation, which is to assume a scale-independent proportionality of the velocity transfer function to the displacement transfer functions
\begin{equation*}
 v = a H(a) \frac{d \ln D_1(a)}{d \ln a} x\,,
\end{equation*}
where $x$ is the particle displacement and $D_1(a)$ is the growing mode total matter linear growth factor\footnote{Sometimes $\frac{d \ln D_1(a)}{d \ln a}$ is further approximated as $\Omega_M(z)^{3/5}$.}. As growth function for individual species is not the total growth function, the Zel'dovich approximation introduces an inaccuracy in the relative velocities of the baryons and CDM. We instead use the perturbation theory variables described above directly. Once the displacement field is generated it is interpolated onto the particle grid using cloud-in-cell.

We do not use second-order Lagrangian perturbation theory \cite{Scoccimarro:1998} as the full species specific velocity and displacement second order transfer functions (including baryon-CDM cross-terms) do not exist in the literature. We defer this to future work.
\subsection{Example Simulations}
\label{sec:simulations}

Our primary test simulations are performed with $2\times 256^3$ particles and box sizes of $300$ Mpc/h (comoving). Initial conditions are generated at $z=99$ and we compare to linear theory power spectra and transfer functions generated using CLASS \cite{CLASS}. We assume a periodic box and a $\Lambda$CDM cosmology with $\Omega_\mathrm{M} = 0.3$, $h = 0.67$, $n_s = 0.97$ and $\sigma_8 = 0.8$. We have set $\Omega_\mathrm{b} = 0.05$ so that $\Omega_{\mathrm{CDM}}/ \Omega_\mathrm{b} = 5$. For the majority of our tests the only force included is Newtonian gravity. Even though we include baryon particles, we generally disable pressure forces. Exceptions are a single simulation to show pressure effects, and our \Lya~forest simulations. When enabled, hydrodynamics is followed using Smoothed Particle Hydrodynamics with a cubic kernel \cite{Springel:2005}. Except for our simulations of the \Lya~forest, cooling and star formation are not included.

Our simulations are performed with the N-body code MP-Gadget and initial conditions were generated with MP-GenIC \cite{yu_feng_2018_1451799}. MP-Gadget is descended from Gadget-2 \cite{Springel:2005}, with heavy modifications for scalability and force accuracy. In particular, as described in Ref.~\cite{Bird:2018}, the force tree is rebuilt completely every timestep, rather than nodes being moved according to the average particle velocity. We have also replaced the force kernel which smoothes the short-range tree force into the long-range particle-mesh force. In Gadget-2 a Gaussian error function is used. For MP-Gadget we instead estimate the smoothing kernel by comparing the particle-mesh force to a simple direct $N^2$ N-body force evaluation, following \cite{Heitmann:2010}.

Our full set of simulations is shown in Table~\ref{tab:simulations}. To demonstrate that we reproduce the results of the literature, \texttt{TWOGRID} is a simulation with species-specific transfer functions and a homogeneous, pre-displacement particle distribution of two offset grids. Grids are offset equally in each Cartesian direction by factors weighted by the contribution of each particle species to the total matter density, so that the total center of mass is unchanged. We have confirmed explicitly, using a small simulation with zero initial displacement, that this is a stable configuration. The \texttt{ADAPTIVE} simulation used the same initial conditions but evolved them using an adaptive gravitational softening set to match the SPH smoothing length.

We performed simulations (\texttt{HALFGLASS}) using the setup described in Section~\ref{sec:glass}, with pre-displacement CDM particles on a regular grid and pre-displacement baryon particles on a Lagrangian glass. To confirm that our results continue to hold on larger scales we performed a simulation (\texttt{BIGGLASS}) with a box of $1$ Gpc/h (a scale large enough that the baryon and CDM transfer functions are approximately equal on scales of the box) and $2\times 768^3$ particles. To check convergence with resolution, we performed a simulation with $300$ Mpc/h and $2\times 512^3$ particles (\texttt{HIRESGLASS}). To confirm the effect of pressure forces, \texttt{HYDROGLASS} uses the same parameters as \texttt{HALFGLASS} but enables hydrodynamics for the baryonic particles.

The \texttt{UNDERSAMP} simulation undersamples the baryon particles so that each CDM and baryon N-body particle has the same mass. For this simulation the pre-displacement CDM and baryon particles were initialized on offset grids, as for the \texttt{TWOGRID} simulation, but the mean inter-particle spacing of the baryon particles was $5$ times wider than the CDM.

Finally, we performed two simulations adapted for the \Lya~forest, in a $120$ Mpc/h box with $2\times 512^3$ particles. These parameters were chosen so that the box was large enough to still include some portion of the BAO, while having sufficient resolution to roughly model the forest. One simulation (\texttt{FOREST}) used species specific transfer functions. A second (\texttt{TOTFOREST}) contains CDM and baryons which are both initialised using the total matter transfer function. Both simulations include radiative cooling following \cite{Katz:1996}, a self-shielding correction from \cite{Rahmati:2013} and a simplified star formation criterion which immediately turns all gas with an over-density $\Delta > 1000$ and a temperature $T < 10^5$ K into stars \cite{Viel:2004}. A uniform meta-galactic ultra-violet background was assumed following Ref.~\cite{Puchwein:2018}.

\begin{table}
\begin{center}
\begin{tabular}{|l|c|c|c|l|}
\hline
Name & Box & $N_\mathrm{CDM}^{1/3}$ & $N_\mathrm{baryon}^{1/3}$ & Comments  \\
\hline
\texttt{TWOGRID}    &   $300$ & $256$ & $256$ & Two offset particle grids. \\
\texttt{ADAPTIVE}    &   $300$ & $256$ & $256$ & As TWOGRID but adaptive gas softenings. \\
\texttt{HALFGLASS}  &   $300$ & $256$ & $256$ & Baryons on a glass, CDM on a grid, see text. \\
\texttt{BIGGLASS}  &   $1000$ & $768$ & $768$ & As HALFGLASS. \\
\texttt{HIRESGLASS}  &   $300$ & $512$ & $512$ & As HALFGLASS. \\
\texttt{HYDROGLASS}  &   $300$ & $256$ & $256$ & As HALFGLASS, but with pressure forces enabled. \\
\texttt{UNDERSAMP}  &   $300$ & $256$ & $150$ & Two offset particle grids, same particle masses. \\
\texttt{FOREST}  &   $120$ & $512$ & $512$ & As HALFGLASS but with \Lya~forest physics. \\
\texttt{TOTFOREST}  &   $120$ & $512$ & $512$ & Total matter transfer functions and \Lya~forest. \\
\hline
\end{tabular}
\end{center}
\caption{Table of simulations. Box size is in comoving Mpc/h. Details are explained in the text}
\label{tab:simulations}
\end{table}

\section{Results}
\label{sec:results}

In this Section we describe the results of our simulations. In Section~\ref{sec:offsetgrid}, we discuss the discrepancy with linear theory. In Section~\ref{sec:halfglass} we show that our preferred setup of a Lagrangian glass for baryonic particles produces good agreement with linear theory. Section~\ref{sec:otherstrat} compares the various simulation strategies which successfully reproduce linear theory.

\subsection{Offset Grid Simulations}
\label{sec:offsetgrid}

\begin{figure}
  \includegraphics[width=0.5\textwidth]{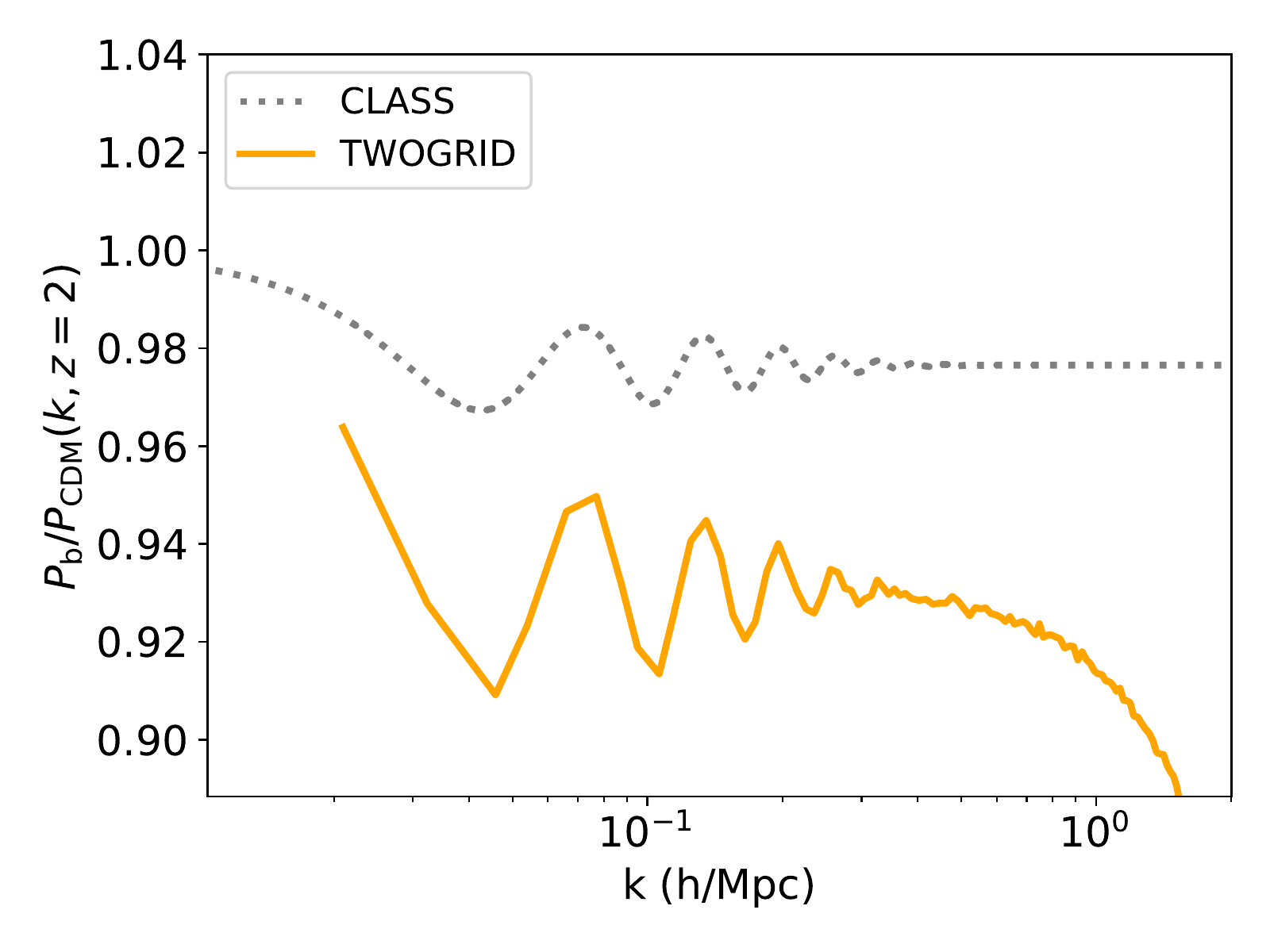}
\includegraphics[width=0.5\textwidth]{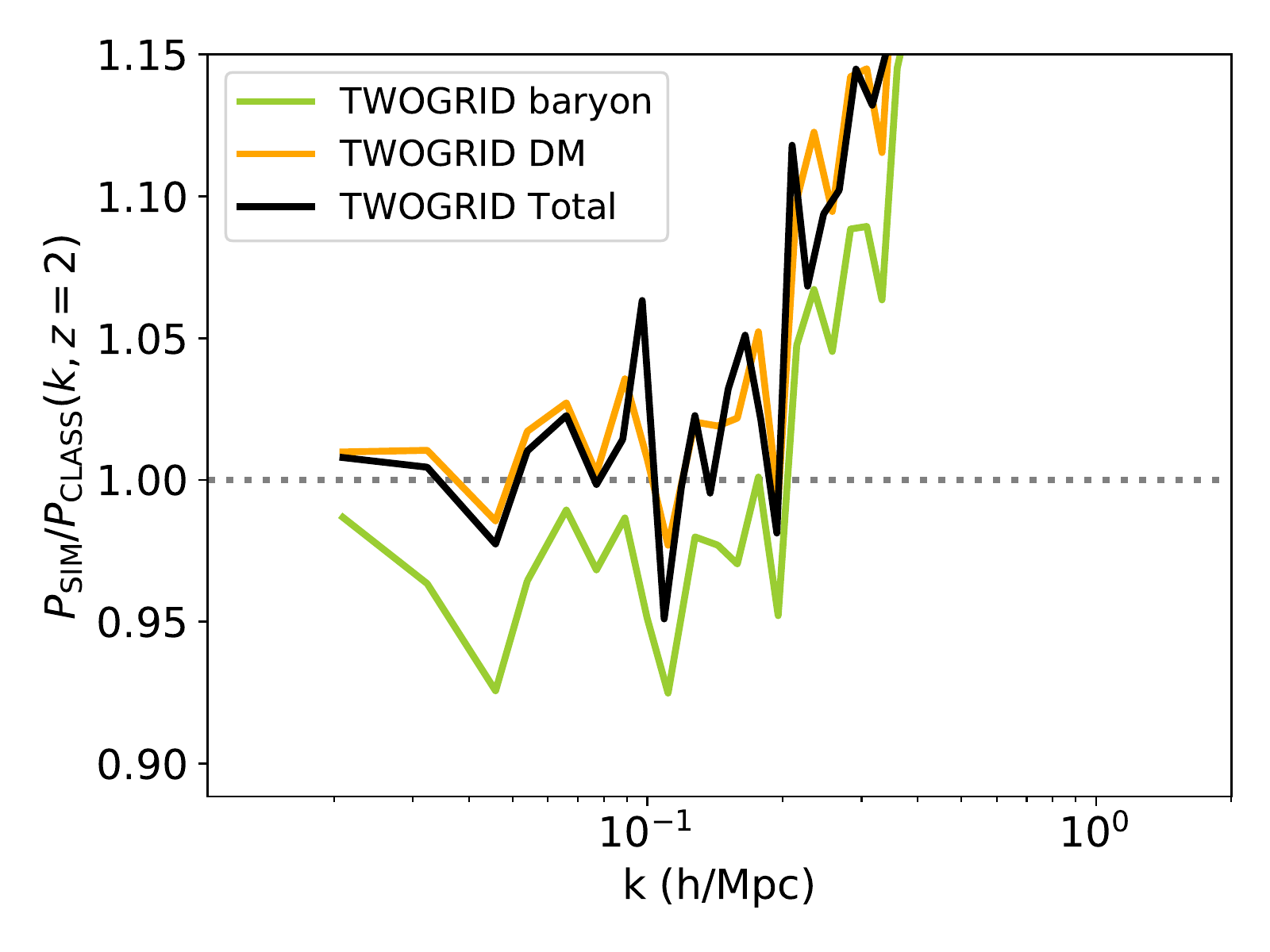} \\
  \includegraphics[width=0.5\textwidth]{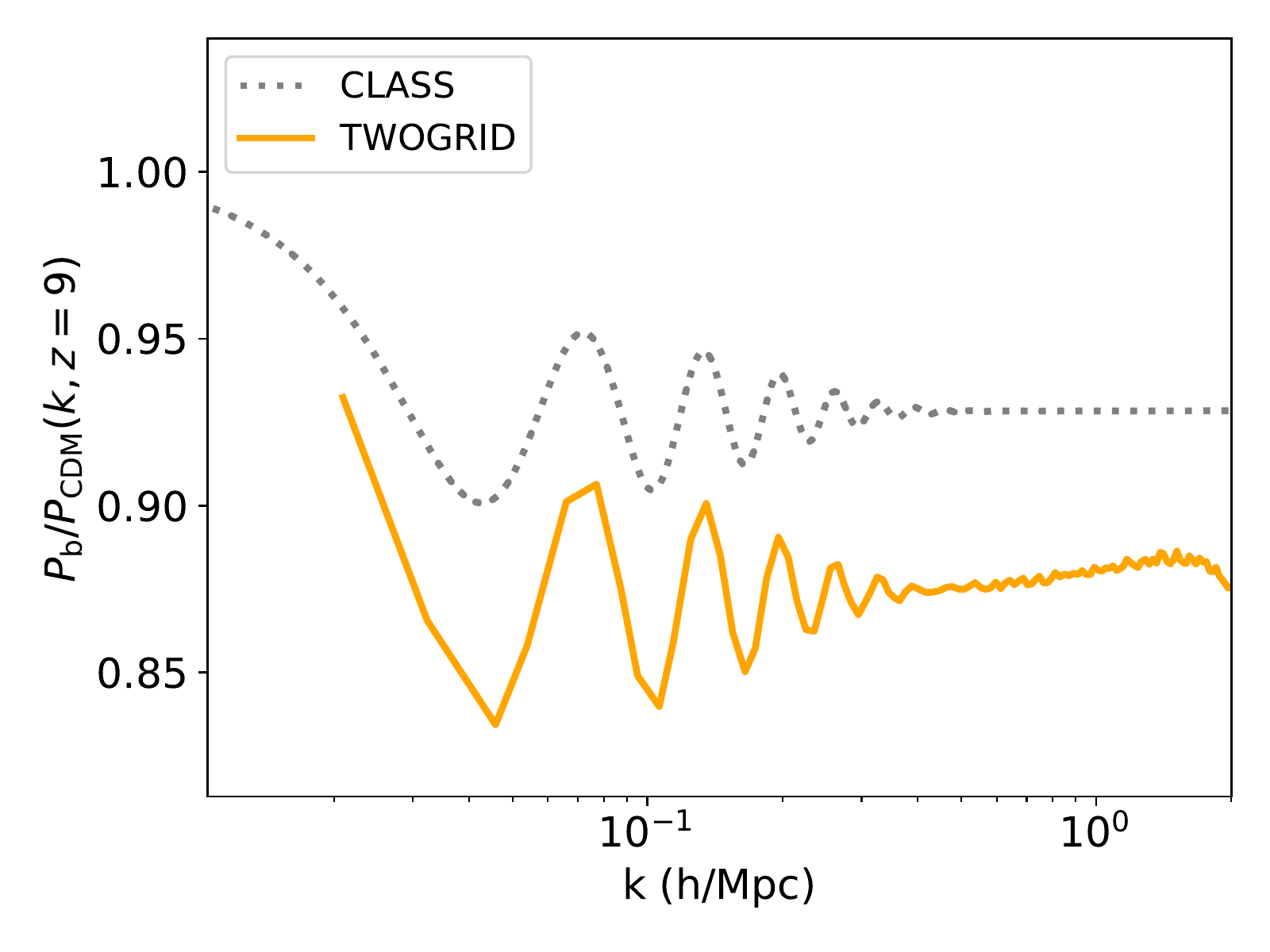}
\includegraphics[width=0.5\textwidth]{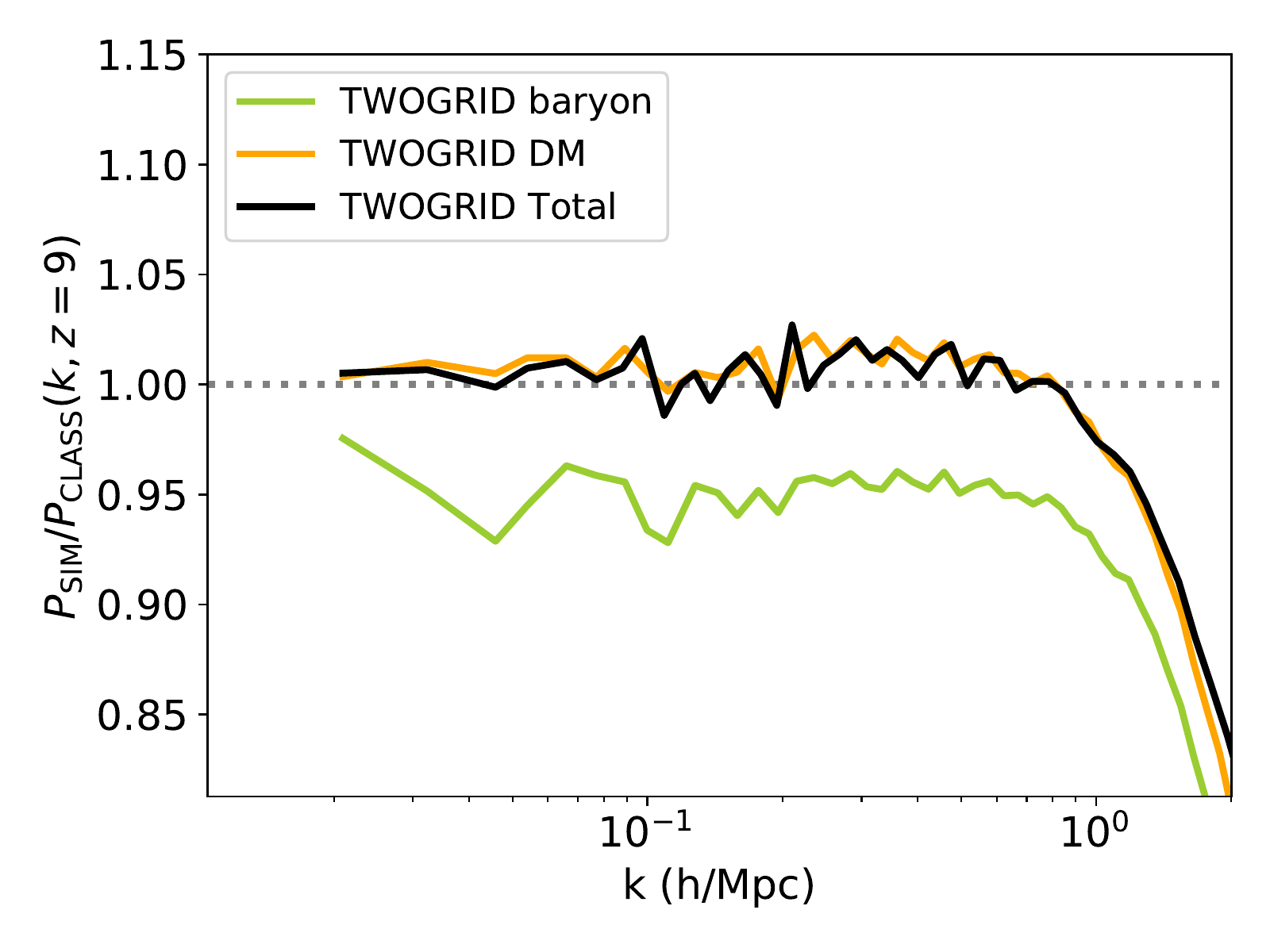}
\caption{Results for the \texttt{TWOGRID} simulation, which initializes particles on two regular grids, offset by half a grid spacing. (Left) Ratio of baryon to CDM power spectra, $P_\mathrm{bar}/P_\mathrm{CDM}(k)$, at $z=2$ for the simulation (solid) and linear theory from CLASS (dot). (Right) Ratio of simulated to linear power spectra, $P_\mathrm{sim}/P_\mathrm{CLASS}(k)$ for gas (green), CDM (yellow) and the total power (black). (Bottom row) At $z=9$.}
  \label{fig:offsetgrids}
\end{figure}

Figure~\ref{fig:offsetgrids} shows baryon and cold dark matter power spectra for the \texttt{TWOGRID} simulation, compared to the output of CLASS at the same redshift. This simulation was initialized using the traditional method of two offset grids of particles. Figure~\ref{fig:offsetgrids} illustrates the problem we aim to solve. Although the total matter power spectrum obeys the linear predictions (right panel), the DM power spectrum is over-predicted and the baryon power spectrum under-predicted, leading to an error in the relative power spectrum.\footnote{The relative power drops at $k > 1$ h/Mpc as $k$ approaches the scale of the particle grid at $5$ h/Mpc. Harmonics of the grid scale begin to be important at the percent level by $k \sim 1$ h/Mpc.}

Disabling the short-range force entirely does allow the simulation to reproduce linear theory \cite{Angulo:2013}, so it is tempting to attribute the discrepancy to a simple force inaccuracy in the Gadget short-range force. We have confirmed that our results are independent of the force accuracy of the simulation. We performed a simulation where the short range force was evaluated using an $N^2$ pairwise gravity solver a simulation where the tree opening angle was changed and a simulation where the split scale between the short and long-range forces was increased. Although the total power in the box in some cases changed on small scales (as expected given the force accuracy), the power ratio between the baryons and cold dark matter was changed by $<0.5\%$. As a further check, we have compared the gravity force from MP-Gadget to an independently written simple $O(N^2)$ particle solver and found good agreement.

We checked that our results were unchanged when all particles had the same timestep and when the long-range PM timestep was $4$ times shorter, ruling out an effect due to adaptive timestepping or the length of the timesteps. We also checked that the results of Figure~\ref{fig:offsetgrids} persist when the box size and particle number were increased. Finally, we have performed the same simulation using the Gadget-2 code, with consistent results, see Appendix \ref{ap:gadget2}. Some of the scales simulated are already non-linear by $z=2$; Figure \ref{fig:offsetgrids} shows a non-linear scale of $k = 0.25$ h/Mpc. There therefore exists the (unlikely) possibility that the discrepancy is a physical effect resulting from non-linear growth. To rule this out, we performed a simulation with suppressed initial power, so that $\sigma_8(z=0) = 0.008$. This simulation gave a similar result to the \texttt{TWOGRID} simulation, but with increased noise on small scales.

Our broad results, that two simulation grids lead to inaccurate power on large scales and the inaccuracy depends on the softening length, match those of Ref.~\cite{Angulo:2013} (AHA13). However, some aspects of our simulations differ. In particular, the simulations of AHA13 produce more power in the baryon component than linear theory predicts. For a similar simulation, we find less power than predicted by linear theory.\footnote{Figure 1 of AHA13 also shows that the magnitude of the discrepancy changes non-monotonically with the smoothing length. For example, the largest $P_{b}/P_\mathrm{cdm}$ occurs with a PM-only code and a smoothing length of $300$ kpc/h, similar to the mean inter-particle spacing of $500$ kpc/h, whereas AHA13 show that a simulation with a softening length of $100$ kpc/h is much closer to linear theory. We performed a similar PM-only simulation using a PM grid cell size of half the mean interparticle spacing and successfully matched linear theory. We have never observed $P_{b}/P_\mathrm{cdm} > 1$ as shown in that figure. We note that if their PM2 and Tree2 simulations were accidentally showing $P_\mathrm{cdm}/P_\mathrm{b}$ then the discrepancy would be monotonic with smoothing length and the PM2 simulation would nearly agree with linear theory and our results. The simulations shown are no longer available, so this is impossible to confirm (Angulo, private communication).} However, the size of the discrepancy found by AHA13 depends strongly on redshift (Angulo, private communication). We therefore suspect some other issue was at play in the earlier work, possibly an implicit choice of a different gauge. Although the physical evolution must be the same on sub-horizon scales for all well-specified gauges, numerical artifacts need not be. Note that the disagreement is due to differences in the initial conditions, not the gravitational evolution, as Appendix~\ref{ap:gadget2} demonstrates that our results are the same when using Gadget-2, the same code as used by AHA13.

\subsection{Combined Glass/Grid Simulations}
\label{sec:halfglass}

\begin{figure}
\includegraphics[width=0.5\textwidth]{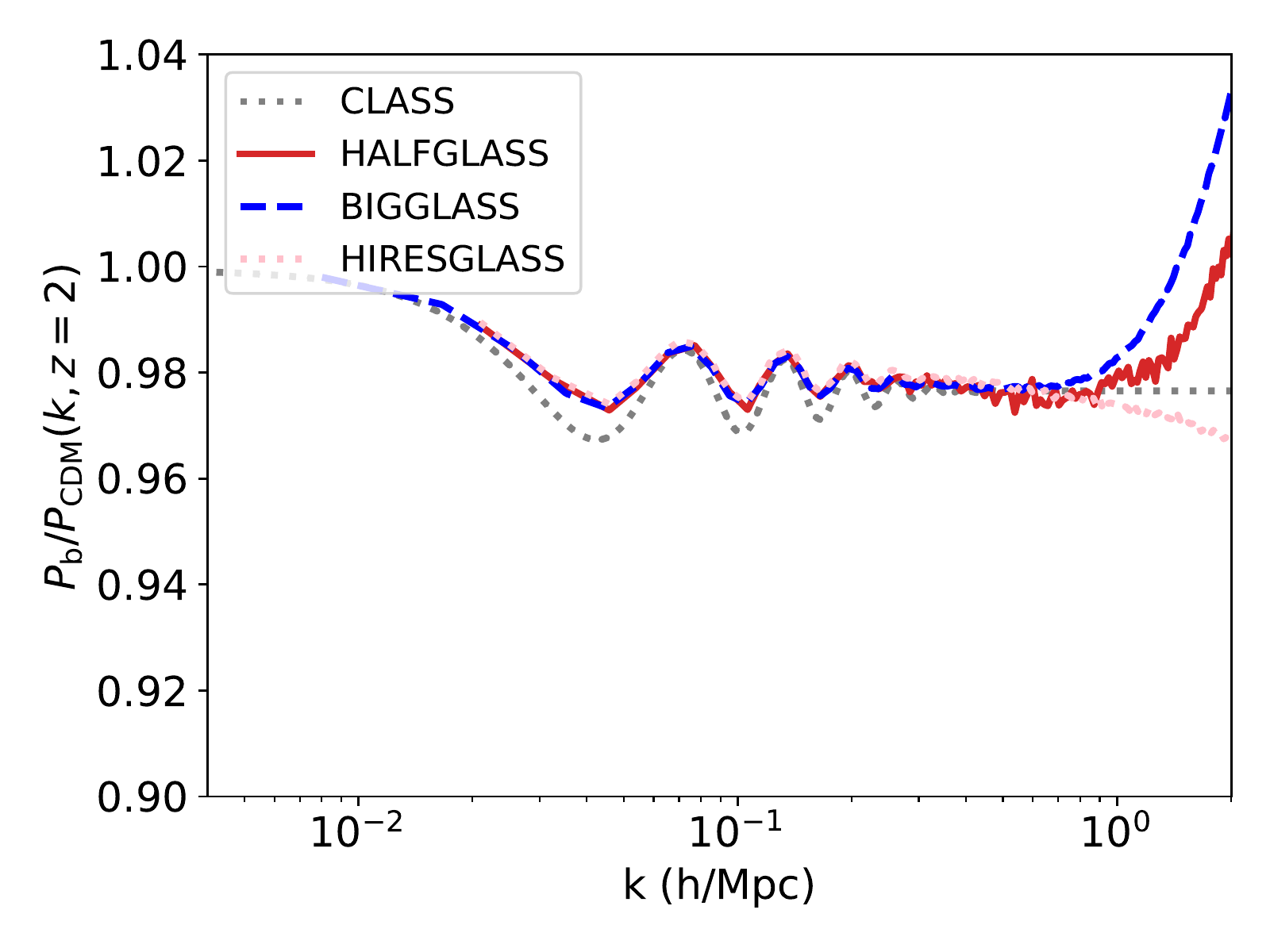}
  \includegraphics[width=0.5\textwidth]{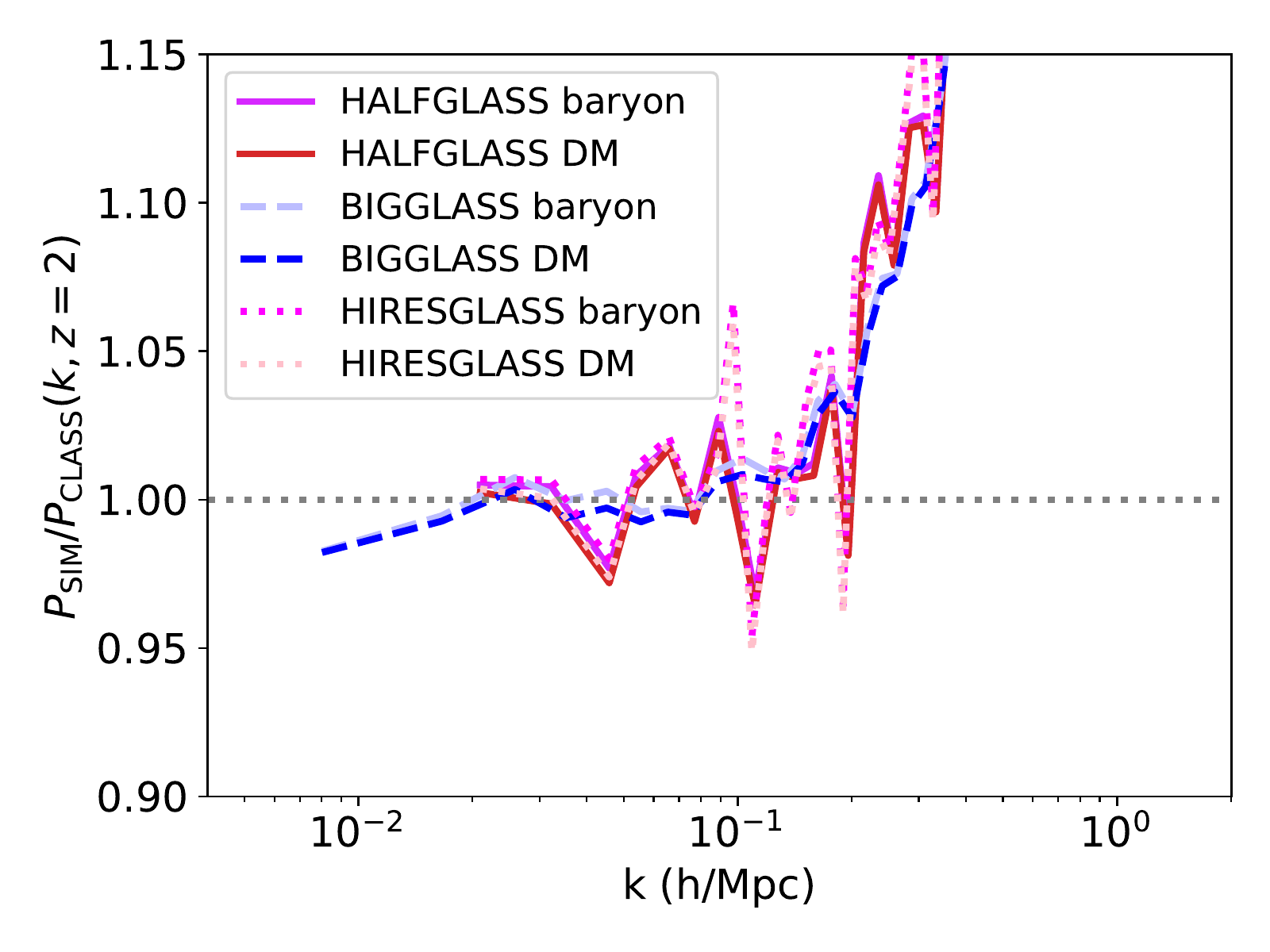}
\includegraphics[width=0.5\textwidth]{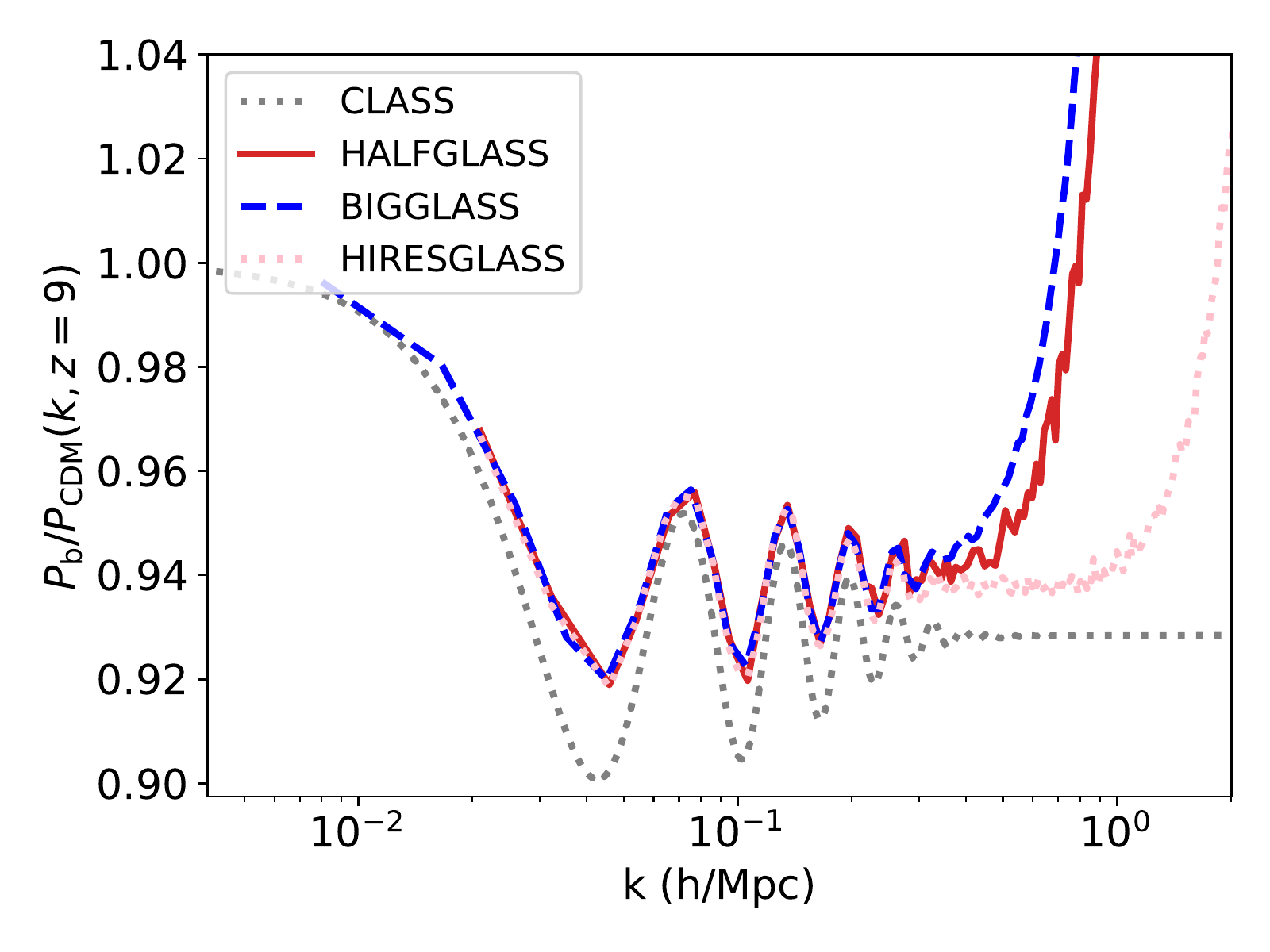}
\includegraphics[width=0.5\textwidth]{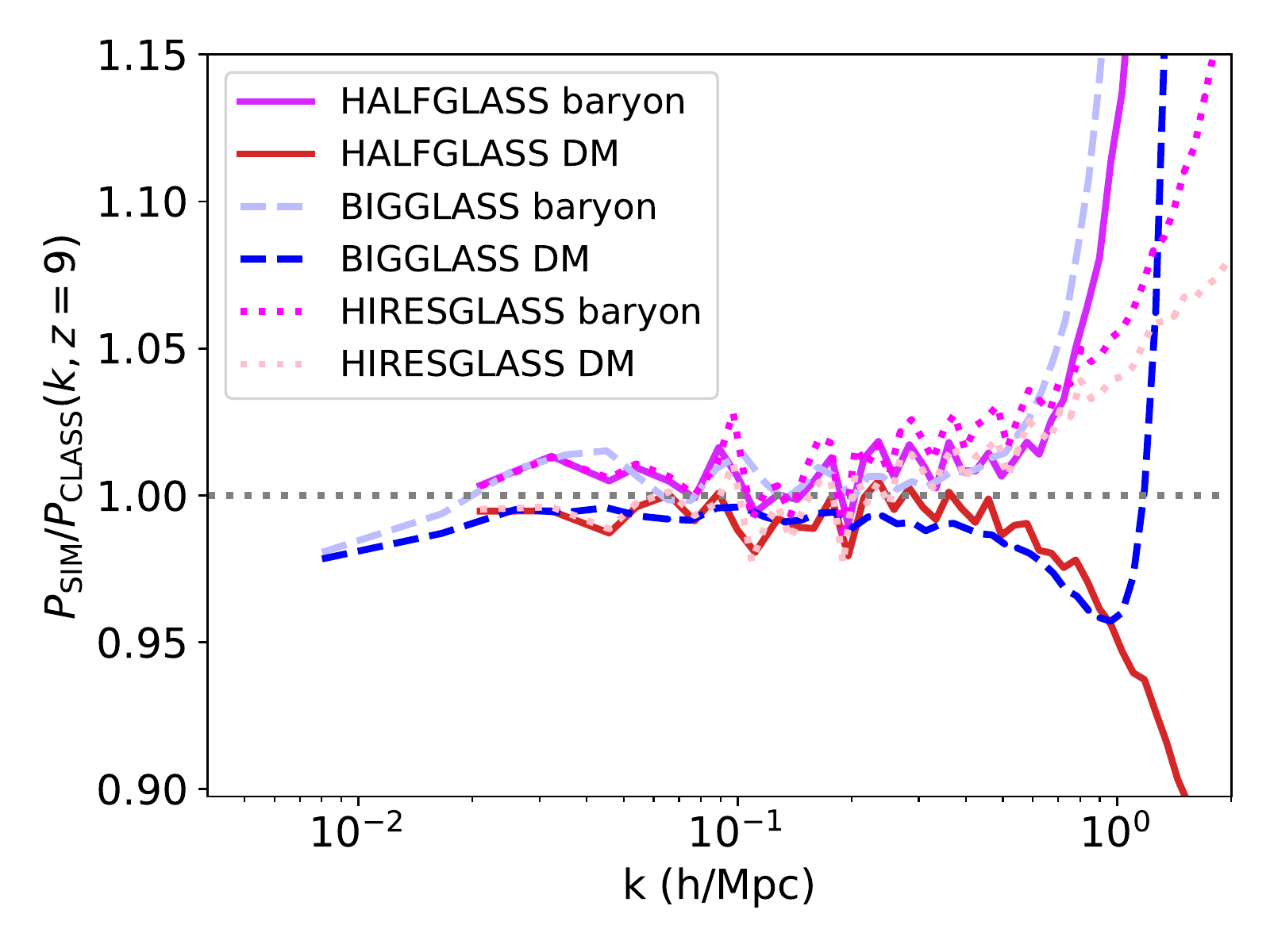}
\caption{Results for our preferred simulation setup, where CDM is initialised with a regular grid and baryons are initialised with a glass. Shown are the \texttt{HALFGLASS} simulation, with $2\times 256^3$ particles in a $300$ Mpc/h box, the \texttt{HIRESGLASS}, with $2\times 512^3$ particles in a $300$ Mpc/h box, and the \texttt{BIGGLASS} simulation, with $2\times 768^3$ particles in a $1000$ Mpc/h box. (Left) Ratio of gas to CDM power spectra, $P_\mathrm{bar}/P_\mathrm{CDM}(k)$, at $z=2$ for the simulations and linear theory from CLASS. (Right) Ratio of simulated to linear power spectra, $P_\mathrm{sim}/P_\mathrm{CLASS}(k)$ for gas and CDM in each simulation. (Bottom row) At $z=9$.}
  \label{fig:baryonglass}
\end{figure}

Figure~\ref{fig:baryonglass} shows the baryon and cold dark matter power spectra for cosmological simulations initialized using a regular grid for the CDM and a glass for the baryons, generated using the procedure described in Section~\ref{sec:glass}. The agreement with linear theory is now good. We show simulations with two separate box sizes, $300$ Mpc/h (\texttt{HALFGLASS}) and $1000$ Mpc/h (\texttt{BIGGLASS}). The ratio between the CDM and baryon power spectra is extremely similar in both simulations, deviating only on very small scales due to the (slightly) lower resolution of the $1000$ Mpc/h simulation.
We also show simulations with the same box sizes but different mean inter-particle spacings, $1.2$ Mpc/h (\texttt{HALFGLASS}) and $0.6$ Mpc/h (\texttt{HIRESGLASS}). The higher resolution simulation produces similar results to \texttt{HALFGLASS} on large scales and indicates that discreteness effects are important on scales of $k=1$ h/Mpc at $z=2$ and $k=0.5$ h/Mpc at $z=9$. The small scale power excess from discreteness noise is a combination of the $k^4$ glass noise and the feature due to the CDM grid, with the latter dominating. A comparison to the CLASS results (right panel) reveals that the absolute values of the power spectra are also reproduced, up to the non-linear scale.

\begin{figure}
\includegraphics[width=0.5\textwidth]{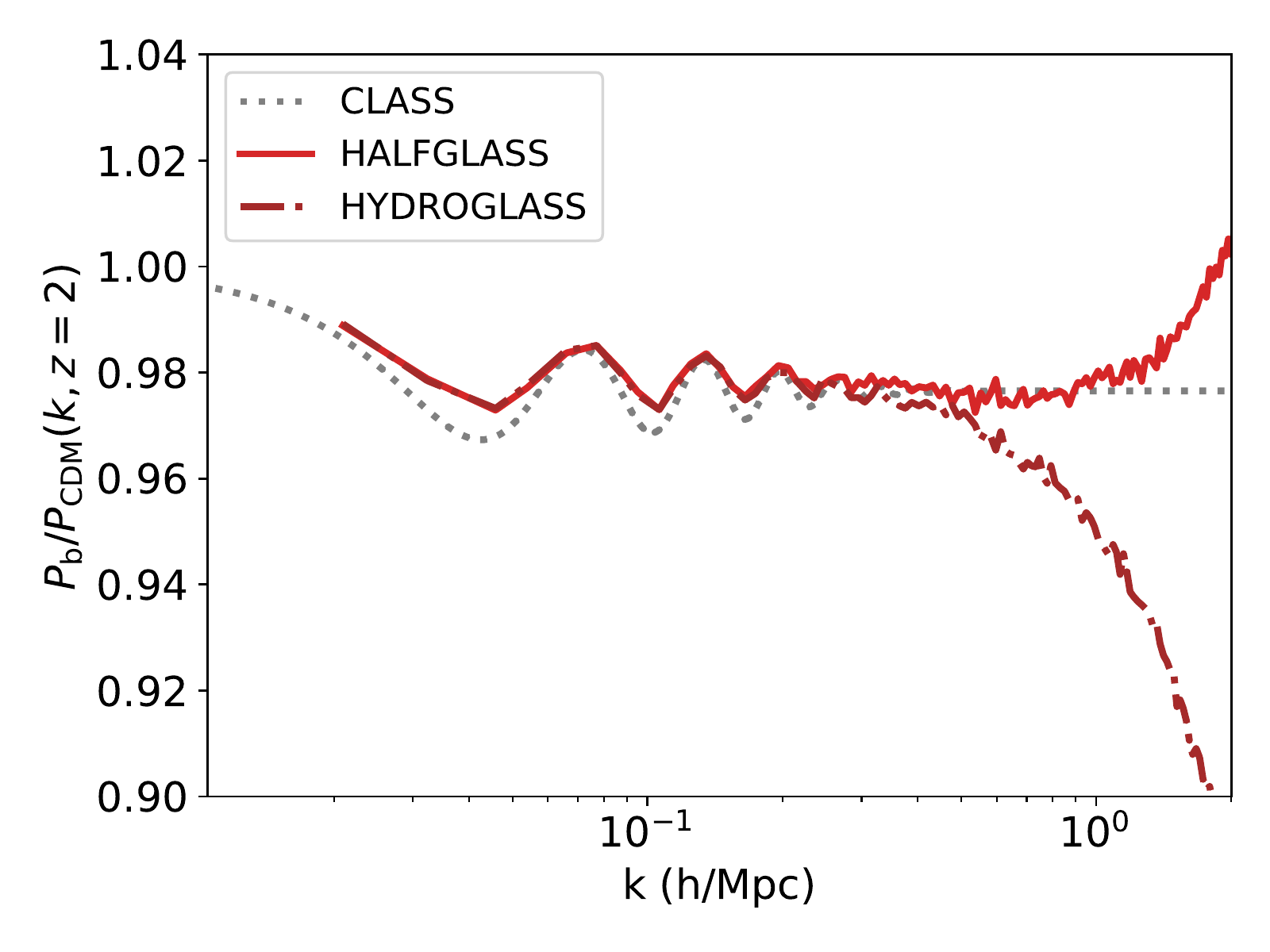}
\includegraphics[width=0.5\textwidth]{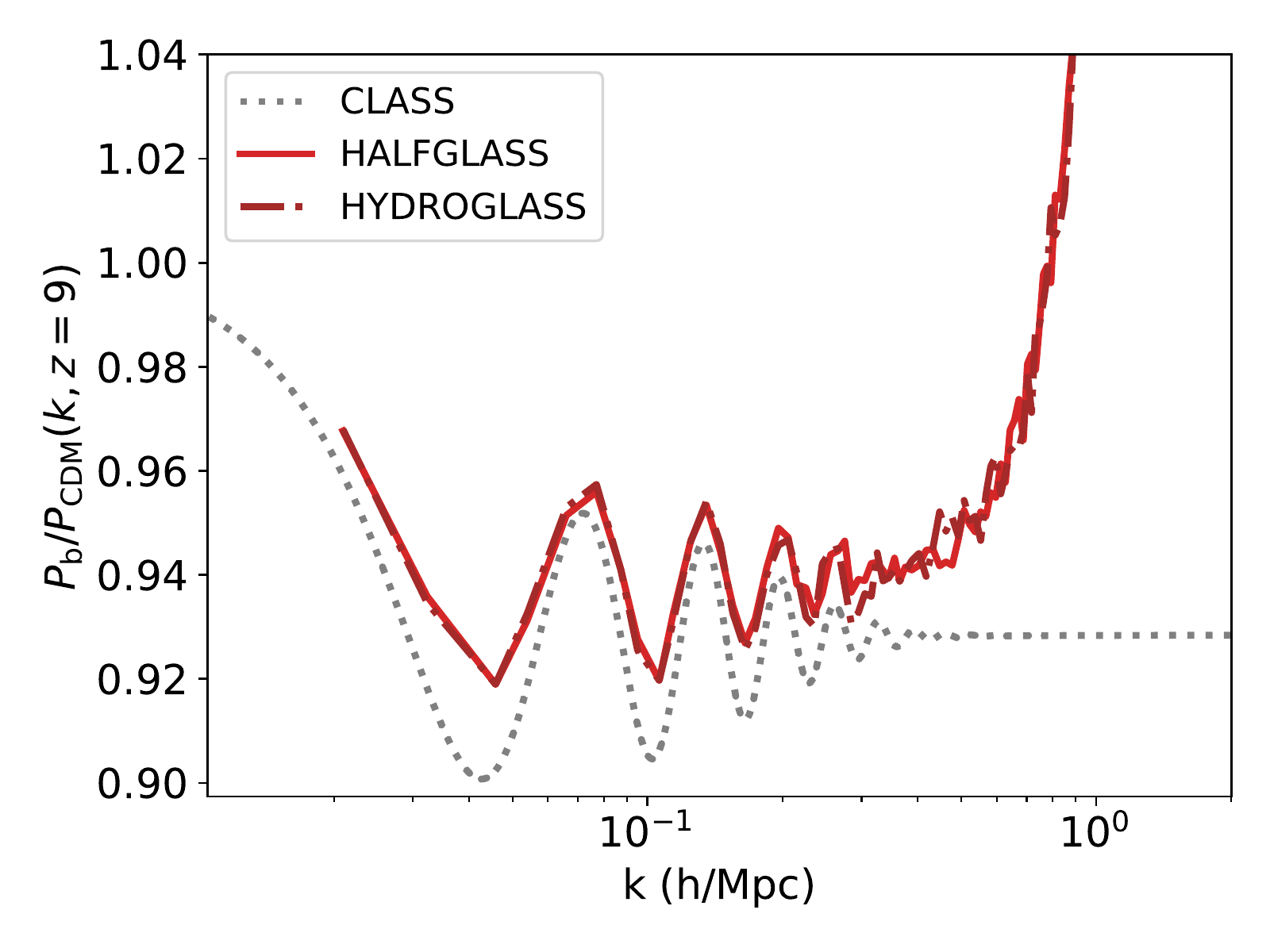}
\caption{Effects of pressure forces. Shown are the \texttt{HALFGLASS} simulation, and \texttt{HYDROGLASS}, which is identical but with pressure forces enabled. (Left) Ratio of gas to CDM power spectra, $P_\mathrm{bar}/P_\mathrm{CDM}(k)$, at $z=2$ for the simulations and linear theory from CLASS. (Right) At $z=9$.}
  \label{fig:hydro}
\end{figure}

Figure~\ref{fig:hydro} shows \texttt{HYDROGLASS}, a simulation similar to \texttt{HALFGLASS}, but with hydrodynamic forces enabled. At $z=2$, the results are similar on scales with $k \lesssim 0.6$ h/Mpc, where the smoothing effect of pressure forces become important. At $z=9$, before reionization, the gas temperature is low and the effect of pressure forces is small. Notice that this shows that in realistic hydrodynamical simulations the $k^4$ glass noise in the baryons will be suppressed.

\subsection{Other Simulation Strategies}
\label{sec:otherstrat}

\begin{figure}
\includegraphics[width=0.5\textwidth]{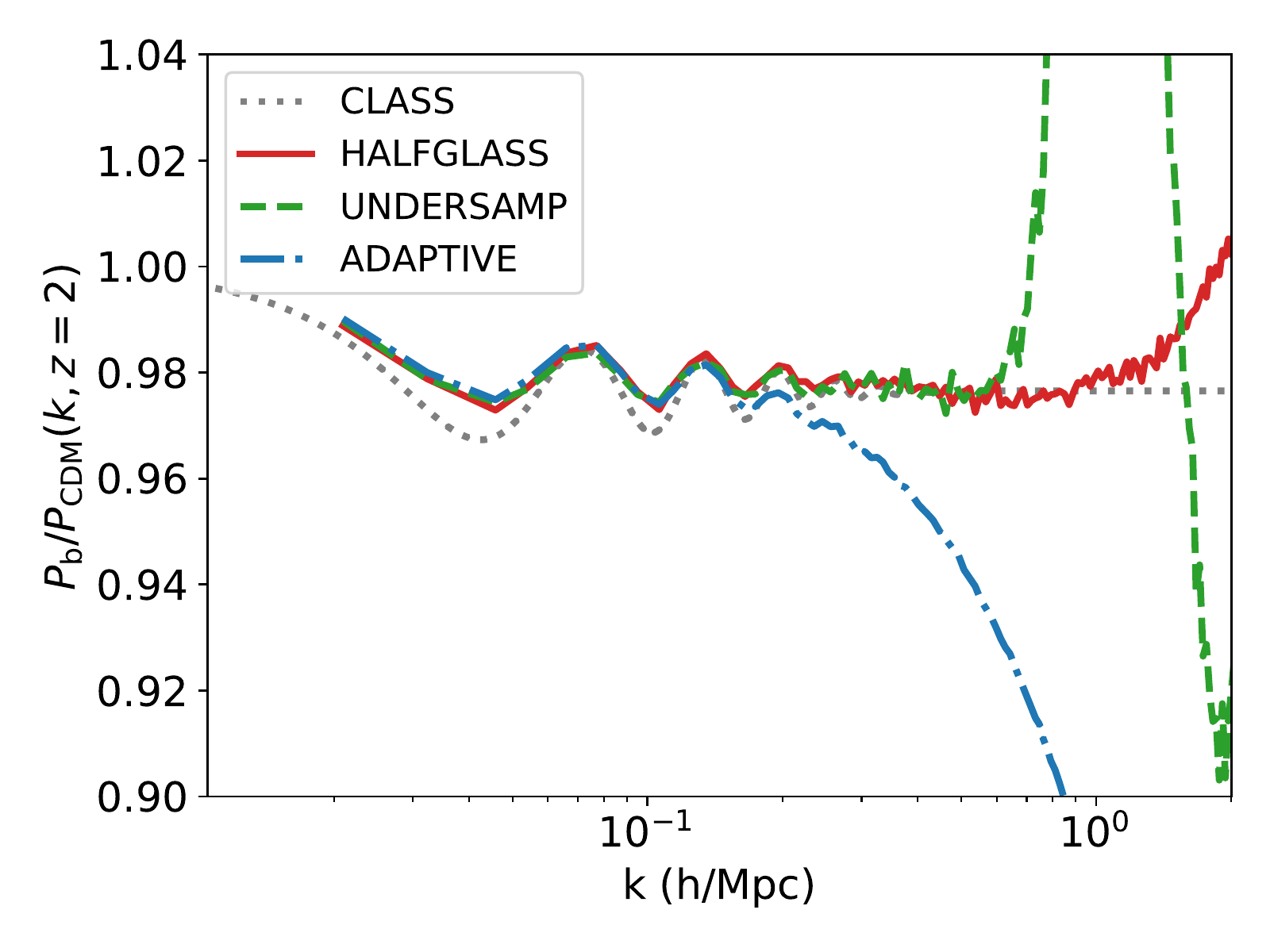}
  \includegraphics[width=0.5\textwidth]{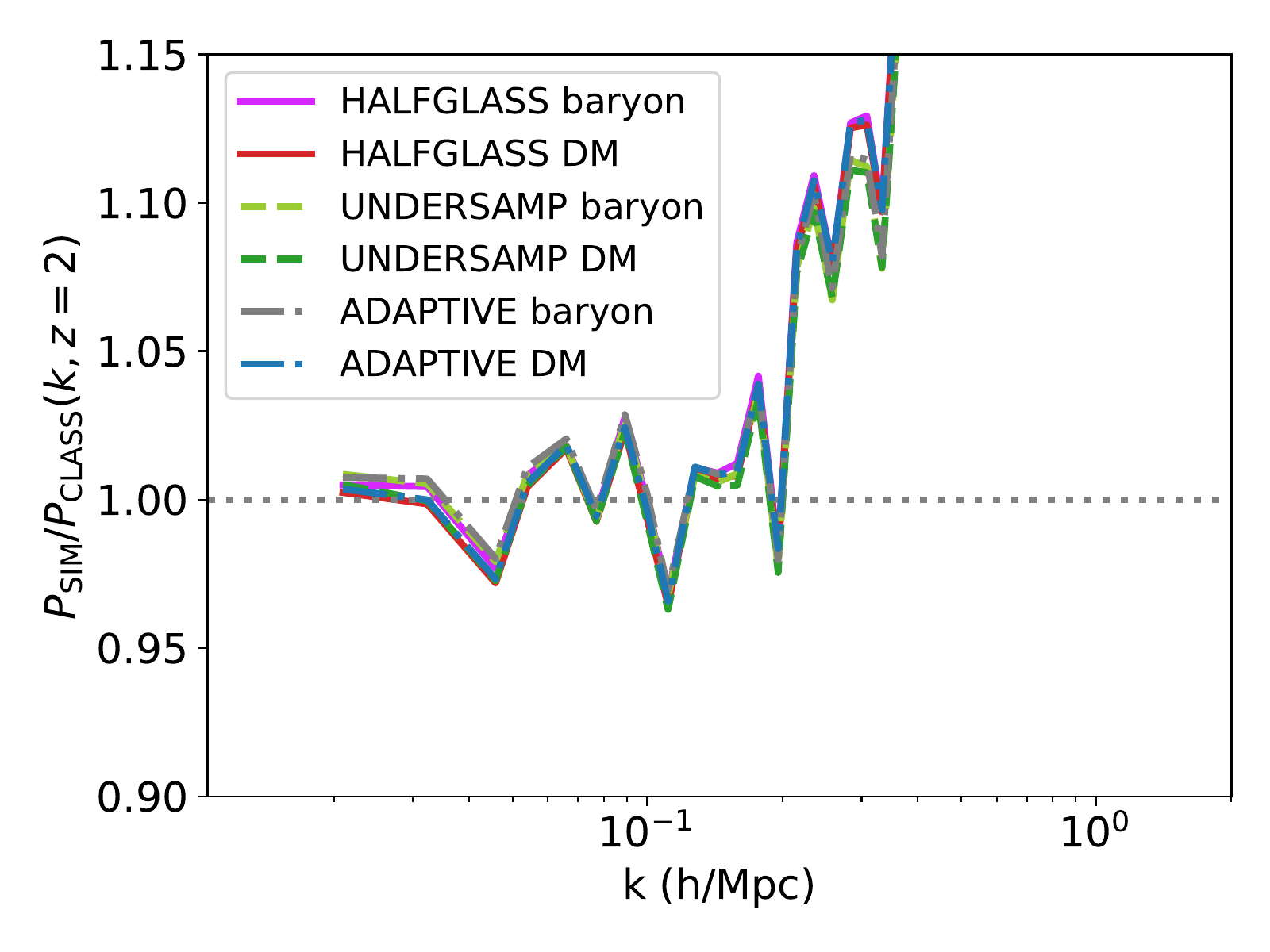} \\
  \includegraphics[width=0.5\textwidth]{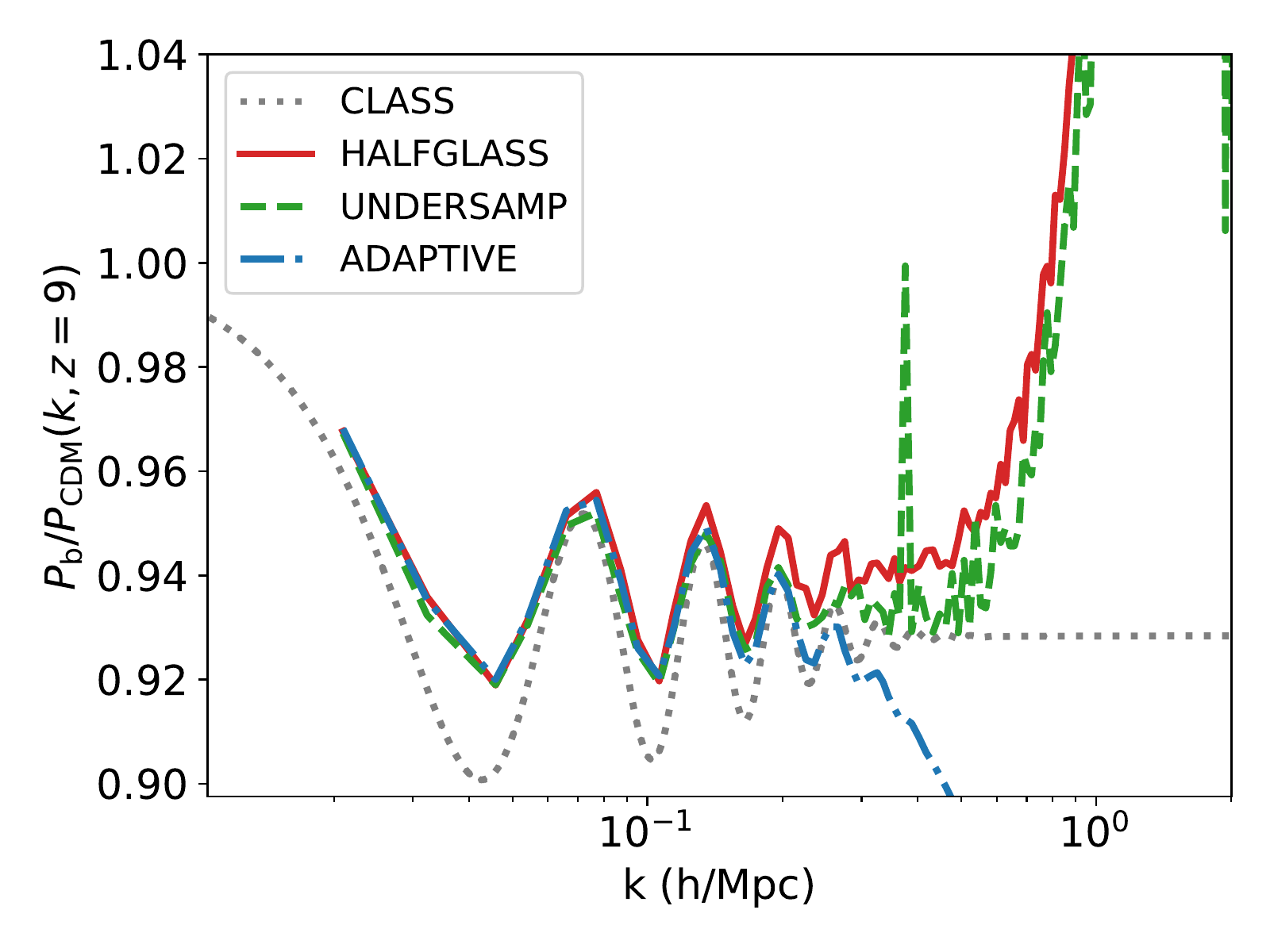}
  \includegraphics[width=0.5\textwidth]{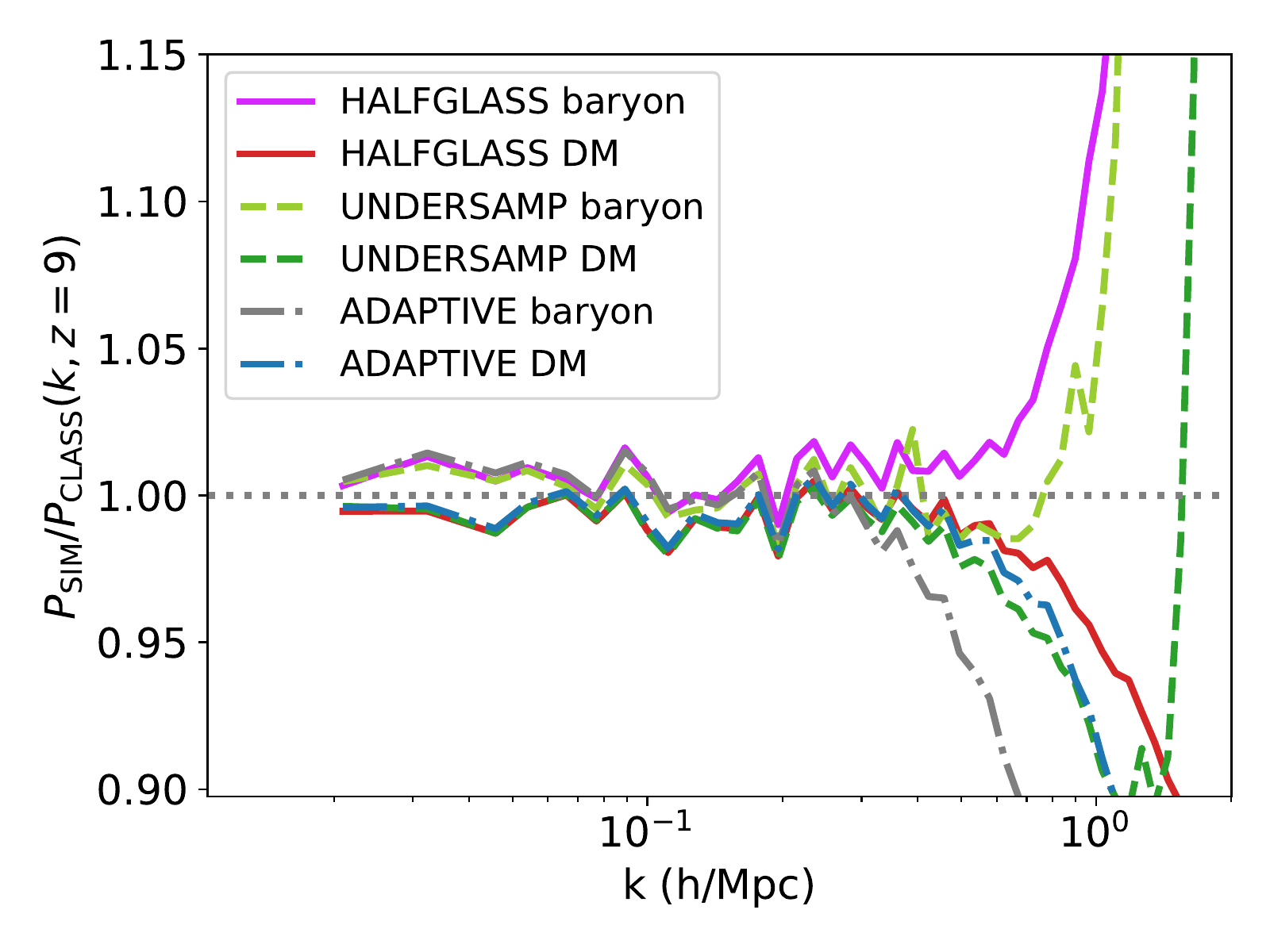}

\caption{Results for other simulation strategies successful at reproducing the relative power of baryons and CDM. Shown are: (\texttt{ADAPTIVE}) initializing particles on two regular grids, offset by half a grid spacing, but with particles evolved using an adaptive softening length for the gas. (\texttt{UNDERSAMP}) Using two regular grids with the number of CDM particle $5$ times larger than the number of baryon particles. Also show is the \texttt{HALFGLASS} simulation, for comparison. (Top row) at $z=2$ (Bottom row) at $z=9$. (Left) Ratio of gas to CDM power spectra, $P_\mathrm{bar}/P_\mathrm{CDM}(k)$, at $z=2$ for the simulations and linear theory from CLASS. (Right) Ratio of simulated to linear power spectra, $P_\mathrm{sim}/P_\mathrm{CLASS}(k)$ for gas and CDM.}
  \label{fig:adaptive}
\end{figure}

Figures~\ref{fig:adaptive} shows the results of other successful simulation strategies at $z=2$. The \texttt{UNDERSAMP} simulation reproduces linear theory as well as the \texttt{HALFGLASS} simulation. However, undersampling of the baryon particles increases their mean separation by a factor of $5$, substantially reducing the effective resolution.
%In addition, undersampling baryon particles would negatively affect more advanced models for star formation and stellar feedback.

The \texttt{ADAPTIVE} simulation, which uses adaptive gravitational softening for baryons, suppresses baryon power by about $2\%$ for $k > 0.3$ h/Mpc at $z=2$. This is the mean effect: the adaptive softening length is smaller inside halos (and thus may be acceptable for galaxy formation, although see \cite{Fvdb:2018}). By comparison, the mean scale at which discreteness effects are important at this redshift is $k = 1$ h/Mpc (see Figure~\ref{fig:baryonglass}). The mechanism by which the \texttt{ADAPTIVE} simulation matches linear theory, smoothing on a scale large enough that the effects of the particle grid are erased, thus substantially reduces the effective simulation resolution as a side-effect.

To further elucidate the effects of our simulations at higher redshift, Figure~\ref{fig:adaptive} (lower row) shows our results at $z=9$. Here the impacts of the initial particle distribution are still directly visible as a small-scale baryon power spectrum rising as $k^4$ and a broadened peak at the particle grid scale for the dark matter. At large scales the effect of the finite number of modes is clearly visible in the sharpness of the peaks. Adaptive softening again suppresses structure formation.

The softening in Gadget affects only the short-range force (at high redshift, when the particle distribution is homogeneous, adaptive gravitational softening is equivalent to disabling the short-range tree force entirely). Adaptive softening is thus capped at the resolution of the particle-mesh grid. For our simulations this is twice the mean inter-particle spacing, so that the maximum scale directly softened is $k \sim 10$ h/Mpc. However, Figure~\ref{fig:adaptive} shows that larger scales are still quantitatively affected, if not suppressed entirely\footnote{In Gadget the gravitational timestep is proportional to the softening and thus adaptive softening also increases the timestep of the baryons. We checked explicitly that this does not affect simulation output.}.

%\footnote{Arguably, if a single particle is of sufficiently low mass to be Jeans stable, baryonic adaptive softenings are the physical thing to do \cite{Fire2:2018}. However, our simulations are of relatively low resolution and our particles are not near this threshold.}.

\section{Discussion}
\label{sec:explanation}

\begin{figure}
\includegraphics[width=0.5\textwidth]{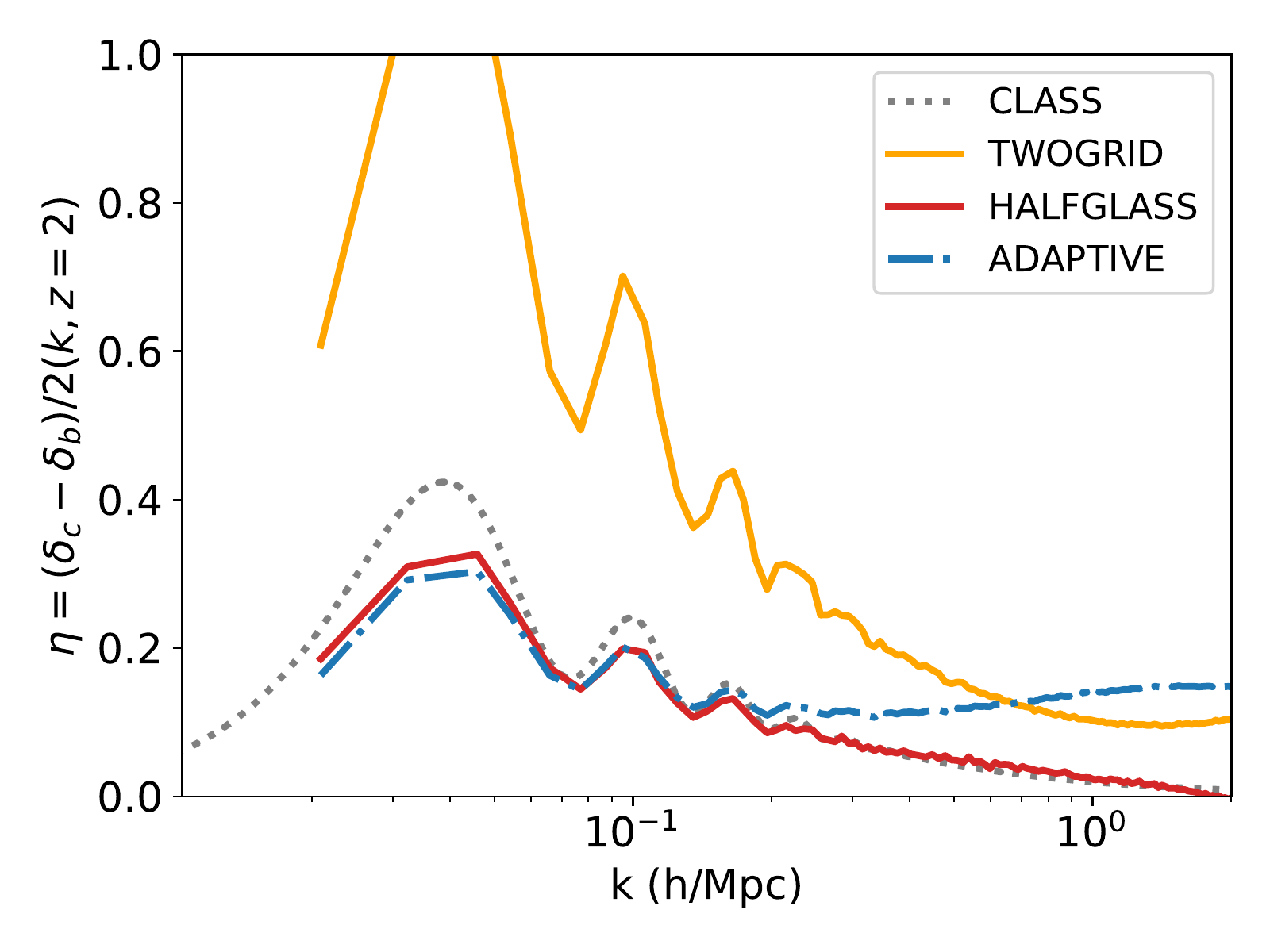}
\includegraphics[width=0.5\textwidth]{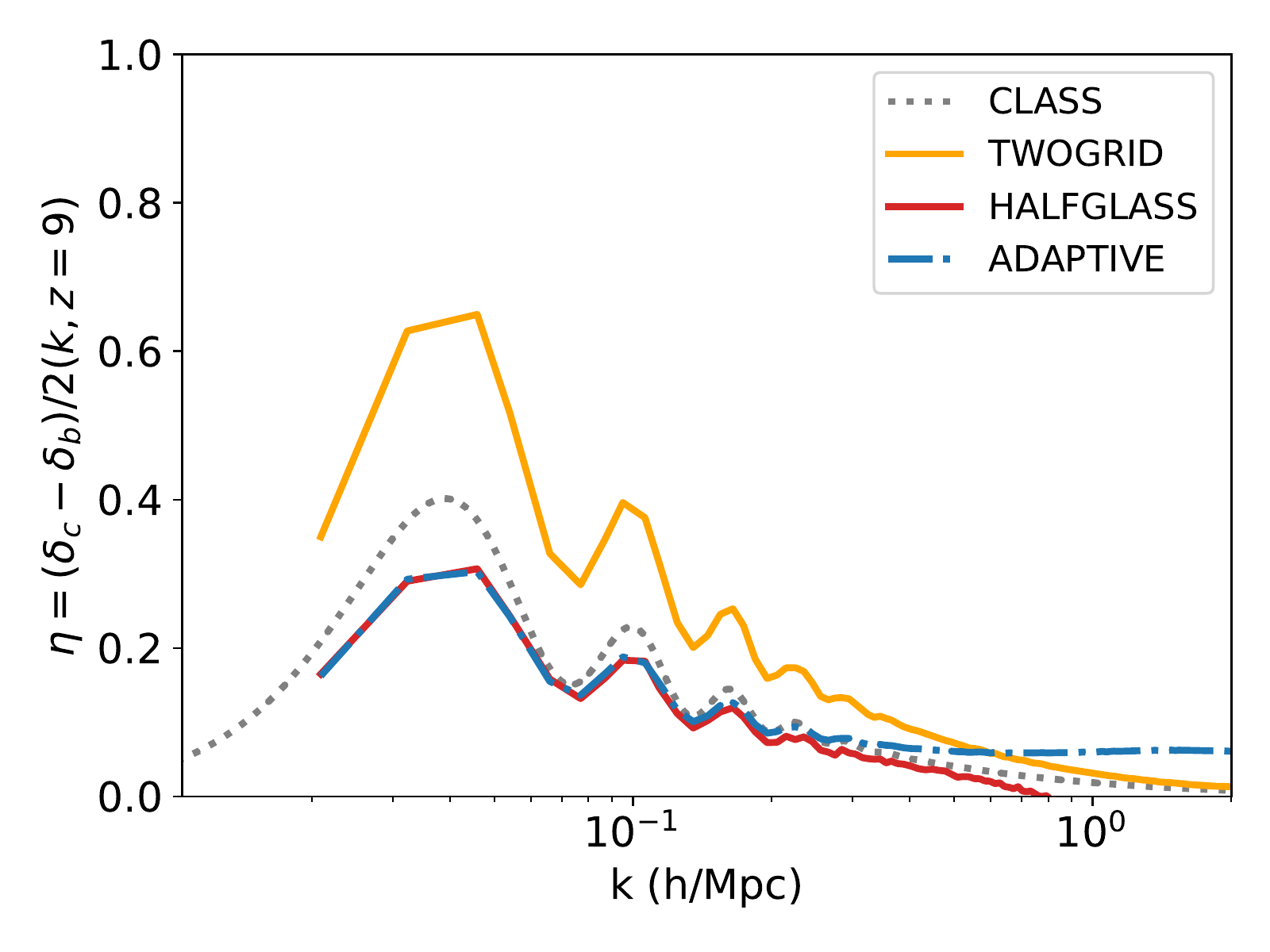}
\caption{The CDM-baryon power difference $\eta$ for the \texttt{TWOGRID} simulation (which initializes particles on two regular grids, offset by half a grid spacing), the \texttt{HALFGLASS} simulation, the \texttt{ADAPTIVE} simulation and from CLASS. (Left) At $z=2$. (Right) At $z=9$.
}
\label{fig:eta}
\end{figure}

To understand these results, we briefly revisit linear cosmological perturbation theory in the synchronous gauge \cite{Ma:1995}. Local energy momentum conservation implies that during matter domination each matter species $X = c$, $b$ obeys
\begin{align}
 \delta_X' + k^2 \theta_X + \frac{1}{2} h' = 0 \\
 \theta_X' + \frac{a'}{a} \theta_X = 0
\end{align}
where a prime denotes a derivative with respect to proper time $\tau$. $\delta_X = \delta \rho_X / \bar{\rho}_X$ is the overdensity, $\nabla^2 \theta_X = \nabla. v_X $ is the velocity divergence and $h$ is the synchronous gauge metric perturbation. We neglect pressure forces and radiation. During matter domination $a \propto \tau^2$.

The growing mode solution is $\delta_X = a \delta_X^0$. By analogy to isocurvature modes, we define $\eta = (\delta_c - \delta_b)/2$, the difference between CDM and baryons. $\eta$ obeys
\begin{align}
 \eta' + k^2 (\theta_c - \theta_b) = 0\,,
\end{align}
which has a decaying solution $\propto a^{-1/2}$ and a constant term sourced during the radiation era. In addition, defining $\xi = (\delta_c + \delta_b) / 2$ we have:
\begin{align}
 \frac{P_b} {P_c} &= \left(\frac{\delta_b} {\delta_c}\right)^2 \\
\frac{\delta_b} {\delta_c} = \frac{\xi - \eta} {\xi + \eta} & = 1 - \frac{ 2 \eta} {\delta_c} = 1 - \frac{ 2 \eta} {a \,\delta_c^0}\,.
\end{align}
Although the linear theory difference in power between baryon and CDM power appears to shrink, it only does so relative to the matter perturbation growing mode.

Figure~\ref{fig:eta} shows the CDM-baryon power difference $\eta$ from the simulations and from linear theory at $z=2$ and $z=9$. The \texttt{HALFGLASS} simulation agrees with linear theory reasonably well on scales not affected by glass discreteness noise ($k < 1$ h/Mpc for $z=2$). \texttt{ADAPTIVE} agrees with linear theory on large scales, but not for $k > 0.3$ h/Mpc, where the baryon power spectrum is suppressed by the increased softening. Simulations not shown in the figure agree with both \texttt{HALFGLASS} and linear theory. However, \texttt{TWOGRID} is discrepant, especially on large scales, indicating the presence of a spurious numerical growing mode in $\eta$. At $z=49$ all simulations agree well with linear theory. However, at lower redshifts $\eta$ grows like $a^{1/2}$ in \texttt{TWOGRID}, dominating as there is no physical growing mode in $\eta$.

While ideally we should understand the origin of this spurious growing mode in \texttt{TWOGRID}, we are unable to produce a convincing analytic explanation and must speculate. The purpose of the regular cubic grid as a particle initialization technique is that the symmetry of the grid suppresses $k^4$ noise in the power spectrum. However, it seems that this symmetry leads to a spurious growth term, perhaps related to the local breaking of isotropy caused by the two grids.
In glass files, the quasi-random positioning of the particles randomizes the small scale relative forces and breaks the symmetry. With adaptive softening all small-scale growing modes are suppressed.

\section{Lyman-Alpha Forest Flux Power Spectrum}
\label{sec:lymanalpha}

% \begin{figure}
% \includegraphics[width=0.5\textwidth]{plots/oversample_2_relpower.pdf}
%   \includegraphics[width=0.5\textwidth]{plots/oversample_2_class.pdf}
% \caption{Results for a $120$ h/Mpc simulation with $512^3$ particles as used for the \Lya~forest. (Left) Ratio of gas to CDM power spectra, $P_\mathrm{bar}/P_\mathrm{CDM}(k)$, at $z=2$ for the simulation (solid) and linear theory from CAMB (dash). (Right) Ratio of simulated to linear power spectra, $P_\mathrm{sim}/P_\mathrm{CLASS}(k)$ for gas (orange) and CDM (blue).}
% \label{fig:lyamatter}
% \end{figure}

\begin{figure}
\includegraphics[width=0.5\textwidth]{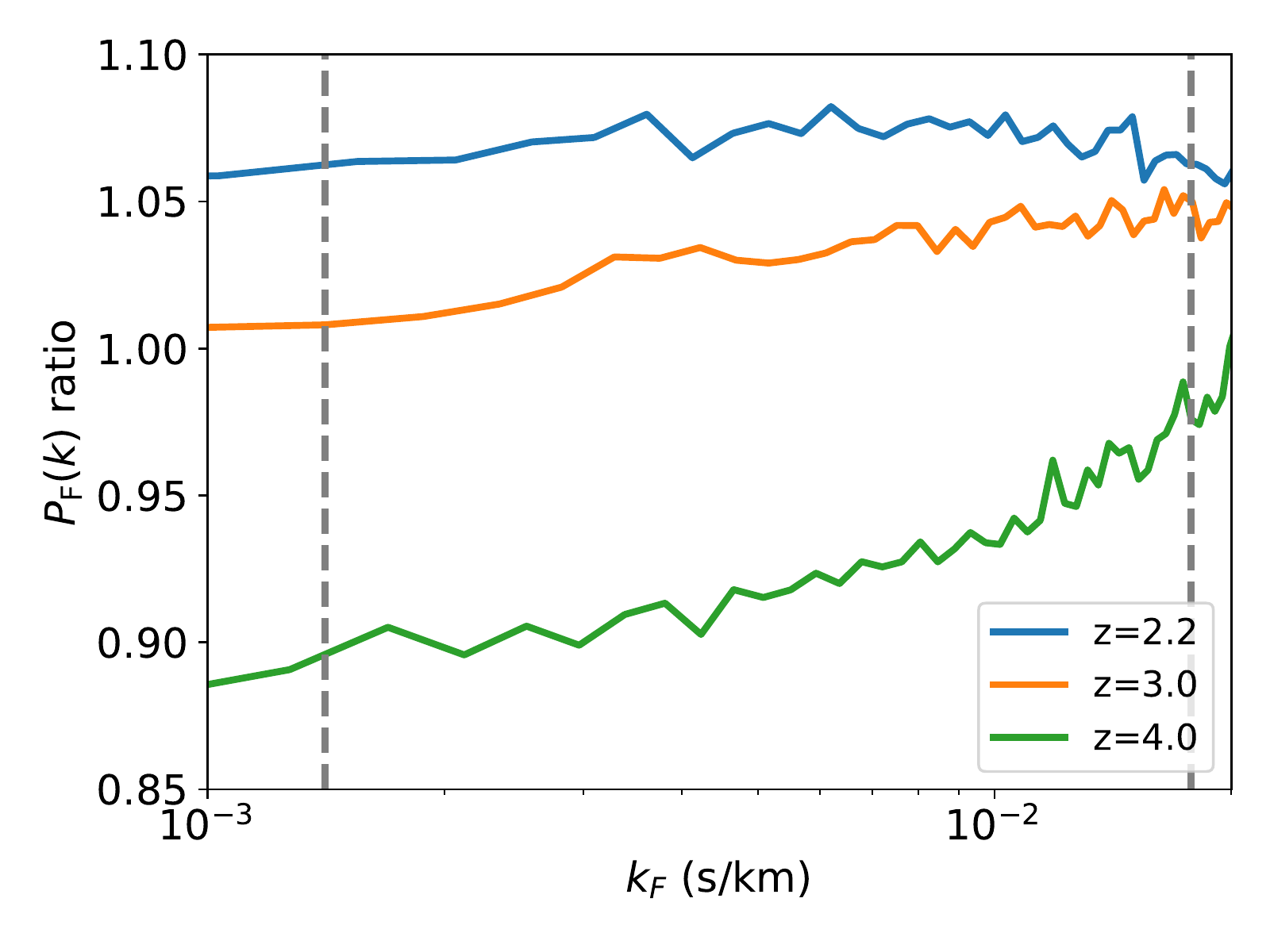}
\includegraphics[width=0.5\textwidth]{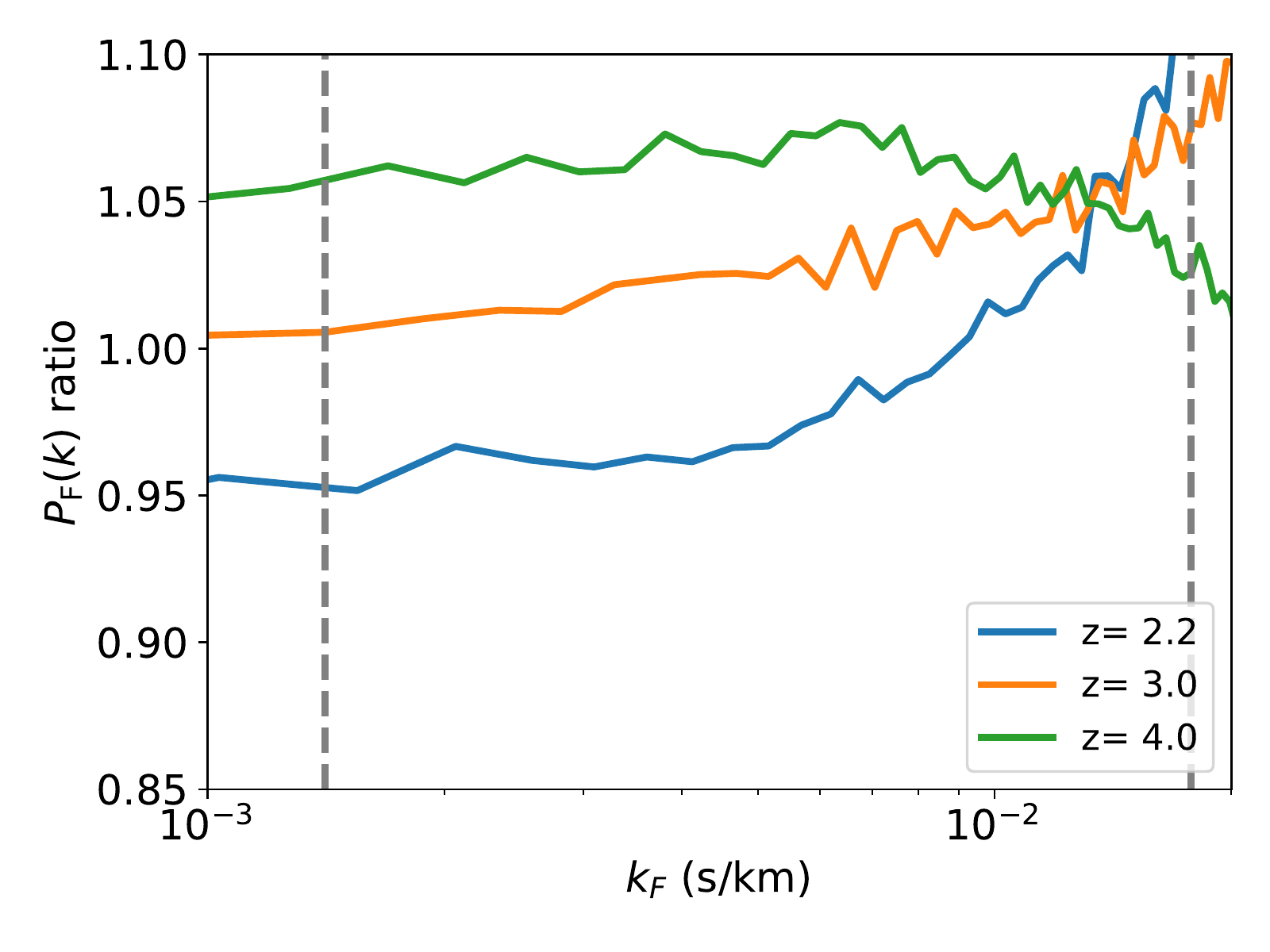}
\caption{Effect of separate transfer functions on the \Lya~forest flux power spectrum. Shown is $P_F(\mathrm{species})/P_F(\mathrm{total})(k)$, so that the reduction in baryon power due to a species-dependent transfer function produces a $P_F(k)\; \mathrm{ratio} < 1$. (Left) The raw change in the flux power between the two simulations. (Right) The change in the flux power after scaling so that each simulation has the same temperature and mean flux. Vertical dashed lines delineate the scales measured by SDSS/BOSS \protect\cite{PD2013}.
}
\label{fig:lyaflux}
\end{figure}

In this Section we evaluate the impact of separate transfer functions on a concrete observable, the \Lya~forest flux power spectrum. We have performed two simulations with $2\times 512^3$ particles and a $120$ h/Mpc box. One simulation (\texttt{FOREST}) uses species specific transfer functions for the baryons and CDM, a grid for the CDM and a glass for the baryons. We have checked that this simulation reproduces $P_\mathrm{bar}/P_\mathrm{CDM}(k)$ equally as well as the \texttt{HALFGLASS} simulation discussed above.
The other simulation (\texttt{TOTFOREST}) uses the total matter power spectrum for both CDM and baryons. Offset grids would still reproduce the linear growth of the total matter power spectrum, and so a simulation using two offset grids with the total matter power spectrum would produce a statistically equivalent result to our preferred glassfile setup\footnote{There would initially be $k^4$ glass noise. However, as the glass distribution is homogeneous, this does not grow and will become dominated by the true power, just like the scale of the CDM particle grid.}. However, to avoid any effect due to a change in the initial realisation of cosmological structure, we opt to use a glass for the baryons even for this simulation. While in previous sections we have used idealised simulations in which the baryons are purely gravitational, here we enable hydrodynamics, cooling and star formation, as described in Section~\ref{sec:simulations}. We generate $32,000$ \Lya~absorption spectra using Ref.~\cite{FSFE}. Sightlines are placed at random positions in the simulation box. The flux power spectrum shown is the 1D power spectrum of the flux, computed along the line of sight to the quasar.

Figure~\ref{fig:lyaflux} shows our results. The left panel shows the ratio in the flux power spectrum between the simulation using species specific initial transfer functions and the simulation using the total matter transfer function. As might be expected, the reduction in the baryonic power spectrum at high redshift leads to a reduction in the flux power spectrum. The effect is larger than that on the matter power spectrum. This may be because the \Lya~forest at high redshift is sensitive to gas at over-density $10-100$, at scales where the dark matter has only recently started to grow non-linearly. Since the power in gas is suppressed, it will start non-linear growth later and the difference between the two species will be increased (as non-linear growth is faster)\footnote{A similar pre-virialization feature is seen in massive neutrino cosmologies \cite{Bird:2012}.}. Ultimately virialization erases differences between gas and dark matter completely.

At $z=2.2$ the power in the \Lya~forest is actually enhanced. This may be because the \Lya~forest is sensitive to redshift space effects. In order for the gas potential to catch up to the dark matter potential, there should be a higher gas velocity power spectrum. Redshift space effects are more important at lower redshifts because the forest probes slightly larger overdensities.

%While the left panel demonstrates that the \Lya~forest is sensitive to the effect of species dependent transfer functions, r
Realistic forest analyses marginalize over the uncertain thermal history of the intergalactic gas as well as the mean flux of the \Lya~forest power spectrum \citep[e.g.][]{PD2013}. The mean flux controls the overall amplitude of the flux power spectrum. The different density distributions in the \texttt{TOTFOREST} and \texttt{FOREST} simulations produce subtly different recombination rates and neutral fractions, and thus have moderately different thermal histories and mean fluxes. In order to establish how much this effect is driving the differences in the flux power spectrum observed in Figure~\ref{fig:lyaflux}, we rescaled each spectrum by a constant factor so that they have the same mean transmitted optical depth, $\bar{\tau} = 0.0023 (1 + z)^{3.65}$ \cite{Kim:2007}. We also fit a power law temperature-density relation and rescaled the temperature of every particle in the box by a density-dependent factor ($\sim 2\%$) so that the slope and intercept of the temperature-density relation were the same in both simulations. The ionization equilibrium neutral fraction was recomputed using the new particle temperatures and new spectra generated.

The right panel of Figure~\ref{fig:lyaflux} shows the flux power spectrum ratios after this rescaling. The effects are now much smaller, on the order of $5\%$. While comparable to the $1\sigma$ statistical errors on the BOSS DR9 1D flux power spectrum, this is larger than the $2-4\%$ statistical error achieved in the newest DR14 data release of Ref.~\cite{Chabanier:2019}.
%Notice that mean flux rescaling reverses the redshift trend, so that \texttt{FOREST} has a larger flux power spectrum at $z=4$ and a smaller one at $z=2.2$.

%Although the effect of a species dependent transfer function is now smaller, it is still present and significant. For comparison, the statistical error on the BOSS 1D \Lya~forest flux power spectrum is a few percent \cite{PD2013}.  after marginalisation of thermal parameters, around $5\%$, there is the potential for this systematic to affect cosmological parameter estimation at the $1\sigma$ level.

We checked the effect of mass resolution on Figure~\ref{fig:lyaflux} using two simulations with $60$ h/Mpc boxes and $2\times 512^3$ particles. The raw flux power spectra ratios shown in the left panel increased by a uniform $5\%$, reflecting a change in the relative mean flux. Once the spectra were rescaled to achieve the same mean flux, the difference was much smaller and consistent with noise for $z >= 3$. At $z=2.2$ the higher resolution simulation exhibited a reduction in power by $\sim 1\%$. For the forest, the effect of finite resolution should be strongest at high redshift \cite{Bolton:2009} and the effect of finite box size strongest at low redshift. We therefore believe that this small change is largely driven by the smaller box and thus Figure~\ref{fig:lyaflux} shows only the lower-resolution, larger box size, simulation.
%Figure~\ref{fig:lyamatter} shows the relative matter power spectra (compared to linear theory) for the box with separate transfer functions, and demonstrates that we still achieve reasonable results for a full hydro simulation.

\section{Conclusions}
\label{sec:conclude}

We have shown that the inability of most cosmological N-body simulations to reproduce the linear theory prediction for the offset between baryons and CDM arises from their use of two offset grids for baryon and CDM particles. We identify a spurious growing mode in this setup proportional to $a^{1/2}$ which dominates for $z \leq 9$.

We demonstrate several possible mechanisms for resolving this issue. Our preferred solution is to use a Lagrangian glass for the baryon particles. Chance close pairs of CDM and baryon particles are mitigated by reversed gravity timesteps run on the total particle distribution. We prefer this solution as it fully resolves the discrepancy with linear theory without compromising the simulation resolution. Other considered mechanisms are adaptive gravitational softenings and under-sampling baryon particles. We show that both reduce the effective resolution of the simulation with fixed particle number. Adaptive gravitational softening in particular means the effective force resolution is reduced to the mean inter-particle spacing.

Finally, we examine whether this linear theory prediction affects a specific observable, the \Lya~forest 1D flux power spectrum. The \Lya~forest is a natural place to look for observational consequences because it is sensitive to neutral hydrogen gas and measures high redshifts.
We find that it does at the $5-10\%$ level. This is larger than the $2-4\%$ statistical error achieved in the current DR14 BOSS data by Ref.~\cite{Chabanier:2019}, and thus has the potential to affect cosmological parameter constraints from that dataset \cite[e.g][]{PD2019}. Given the complexity of the \Lya~likelihood function, the exact parameters affected and the magnitude of the shift is not clear, and we defer this question to future work using a larger simulation suite.

We have not investigated the impact on small-scale constraints on the warm dark matter mass \cite{Irsic:2017} as our current simulations do not resolve the flux power spectrum on scales of $k = 0.01 - 0.1$ s/km ($k = 0.6 - 6$ h/Mpc at $z = 4$). The statistical errors on these scales are still of order $10\%$, so it is likely that the systematic we discuss here has minimal effect. However, our work demonstrates that a species dependent transfer function could be important for future constraints. There may also be observational implications for future high redshift $21$cm or intensity mapping constraints.

% We have shown that the spurious coupling present in N-body simulations between unequal mass CDM and baryon particles can be prevented from affecting the large scale power by setting the initial distribution of baryon particles with a Lagrangian glass. The CDM is initially placed on a grid. Chance juxtapositions of CDM and baryons are avoided by evolving the combined particle distribution with a reversed gravitational force, as in glass generation. This procedure allows a simulation to reproduce linear expectations for the power spectrum ratio between different particle species while avoiding the suppression of small scale power that results from increasing the gravitational softening length. We also explain an alternative possibility: over-sampling the CDM particles so that each particle species has the same mass, and demonstrate that this also reproduces linear expectations for the power spectrum ratio between different particle species.

\acknowledgments

We thank Raul Angulo, Pat McDonald, Tom Kitching, Yin Li, Matt McQuinn, Jose Onorbe, Andrew Pontzen, Martin Rey, Francisco Villaescusa-Navarro and Matias Zaldarriaga for helpful discussions and the anonymous referee for insightful and helpful comments. SB was supported by NSF grant AST-1817256. Computing resources were provided by NSF XSEDE allocation AST180058. This work was partially funded by the UCL Cosmoparticle Initiative. This research was partially supported by the Munich Institute for Astro- and Particle Physics (MIAPP) which is funded by the Deutsche Forschungsgemeinschaft (DFG, German Research Foundation) under Germany´s Excellence Strategy – EXC-2094 – 390783311.

\appendix

\section{Gadget-2 Results}
\label{ap:gadget2}

\begin{figure}
\includegraphics[width=0.5\textwidth]{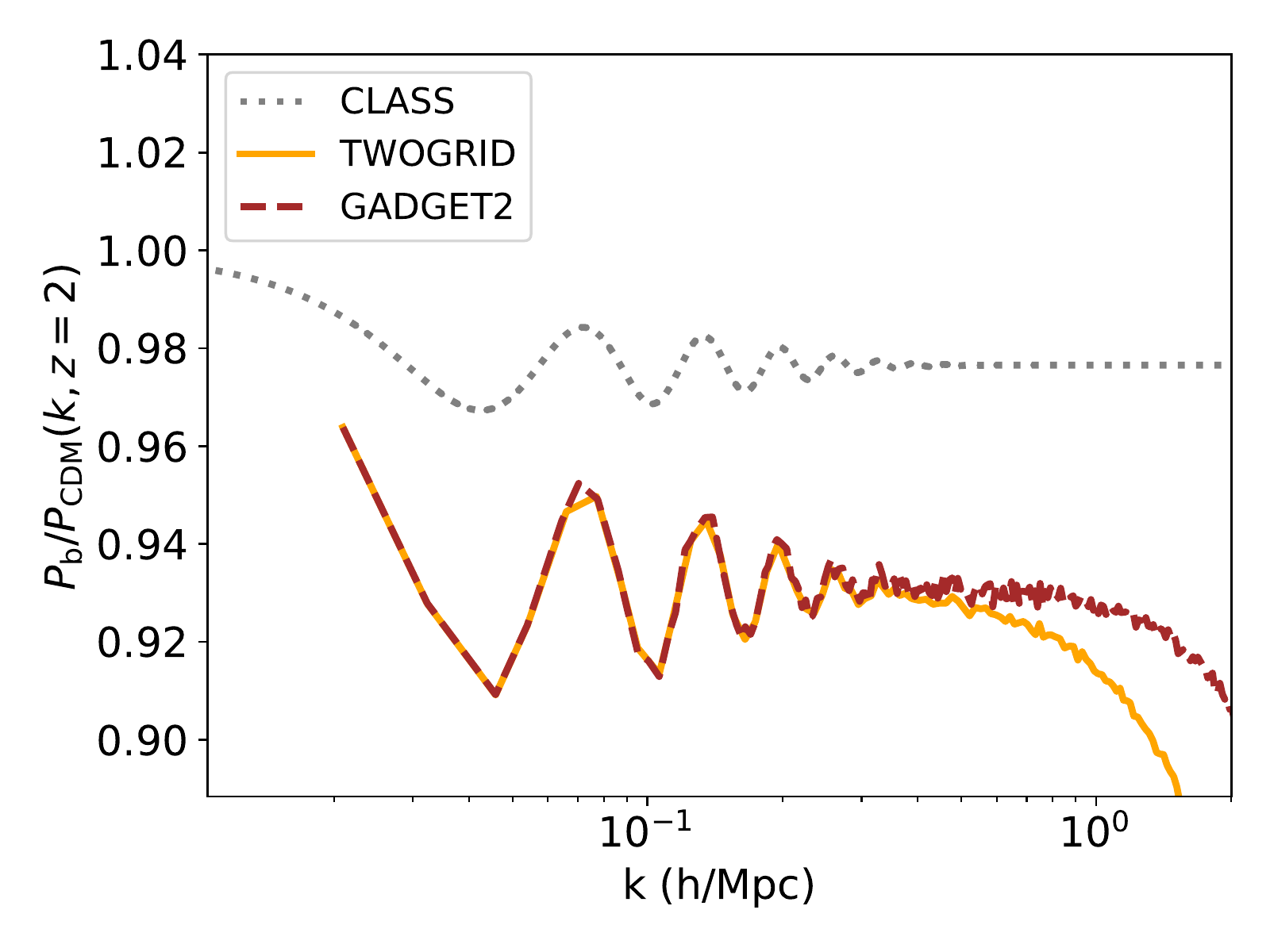}
\includegraphics[width=0.5\textwidth]{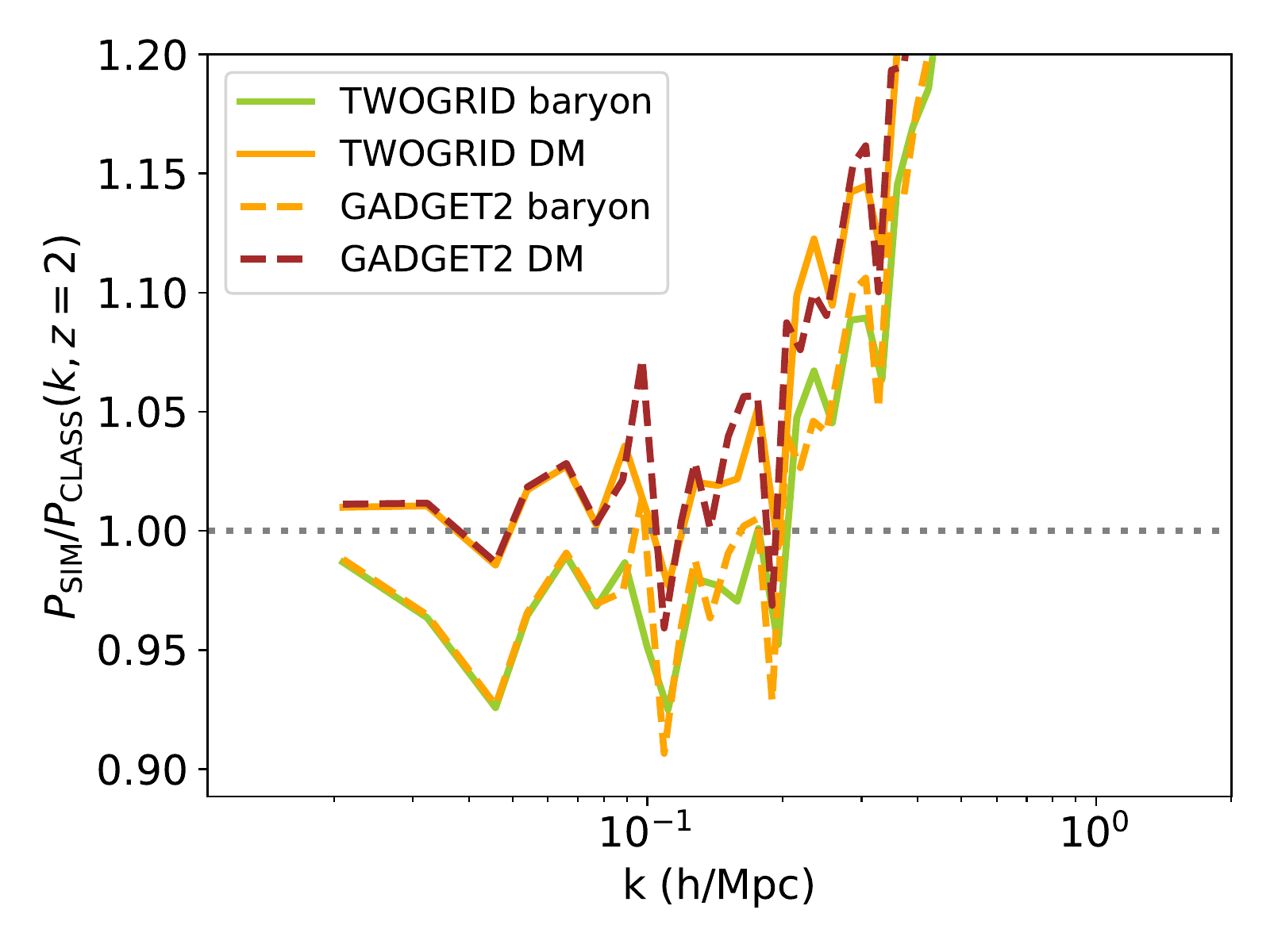}
\caption{Results for a simulation using the Gadget-2 code, compared to an identical simulation using MP-Gadget. Results are consistent, although notice the increased noise in the Gadget-2 force evaluation. (Left) Ratio of baryon to CDM power spectra, $P_\mathrm{bar}/P_\mathrm{CDM}(k)$, at $z=2$ for the simulations and linear theory from CLASS. (Right) Ratio of simulated to linear power spectra, $P_\mathrm{sim}/P_\mathrm{CLASS}(k)$ for gas (green), CDM (yellow) and the total power (black).}
  \label{fig:gadget2}
\end{figure}

As a final check that our results are not due to a bug introduced in MP-Gadget we have reproduced them using the public release of Gadget-2, version 2.0.7. To generate these results we took the initial conditions for our \texttt{TWOGRID} simulation, and converted them to Gadget-2's HDF5 format. Gadget-2 was patched to include radiation in the cosmological background evolution and to disable hydrodynamic pressure forces. All gravitational accuracy parameters were left at their default values. The particle-mesh grid was $512^3$ cells and a softening length of $1/30 \times$ the mean interparticle spacing was used, matching our other simulations. After reaching $z=2$, the snapshots were converted back to MP-Gadget's BigFile format for power spectrum generation.

Figure~\ref{fig:gadget2} shows the results. There is extremely good agreement between MP-Gadget and Gadget-2 on large scales.  Higher redshift snapshots show similar behaviour. On small, non-linear, scales, there is $2\%$ more power in the baryon component in Gadget-2. This is due to increased noise (and thus power) in Gadget-2 at high redshift, which excessively broadens the transient due to the grid scale. The improved short-range smoothing kernel in MP-Gadget suppresses force evaluation noise in highly homogeneous environments.

\bibliographystyle{JHEP}
\bibliography{offset}
\end{document}